\documentclass[12pt]{article}
\usepackage{amsmath}
\usepackage{graphicx,psfrag,epsf, graphics}
\usepackage{enumerate}
\pdfoutput=1
\usepackage[square,sort,nonamebreak,comma]{natbib}
\usepackage{float}
\usepackage{bm}
\usepackage{hyperref}
\usepackage{dsfont}

\usepackage{algorithm}
\usepackage[noend]{algpseudocode}

\makeatletter
\def\BState{\State\hskip-\ALG@thistlm}
\makeatother

\usepackage{mathtools} 
\usepackage{scalerel}
\newcommand{\widesim}[2][1.5]{
  \mathrel{\overset{#2}{\scalebox{#1}[1]{$\sim$}}}}

\newcommand{\blind}{0}

\begin{document}

\def\spacingset#1{\renewcommand{\baselinestretch}%
{#1}\small\normalsize} \spacingset{1}


\if0\blind
{
  \title{\bf Assessing Dynamic Effects on a Bayesian Matrix-Variate Dynamic Linear Model: an Application to fMRI Data Analysis.}
  \author{Johnatan Cardona Jim\' enez \hspace{.2cm}\\
    Institute of Mathematics and Statistics, University of S\~ao Paulo, Brazil\\
   Carlos A. de B. Pereira\\
        Institute of Mathematics and Statistics, University of S\~ao Paulo, Brazil\\
Victor Fossaluza\\
       Institute of Mathematics and Statistics, University of S\~ao Paulo, Brazil    }
  \maketitle
} \fi

\if1\blind
{
  \bigskip
  \bigskip
  \bigskip
  \begin{center}
    {\LARGE\bf Title}
\end{center}
  \medskip
} \fi

\bigskip
\begin{abstract}
In this work, we propose a modeling procedure for fMRI data analysis using a Bayesian Matrix-Variate Dynamic Linear Model (MVDLM). With this type of model, less complex than the more traditional temporal-spatial models, we are able to take into account the temporal and -at least locally- the spatial structures that are usually present in this type of data. Despite employing a voxel-wise approach, every voxel in the brain is jointly modeled with its nearest neighbors, which are defined through a euclidian metric. MVDLM's have been widely used in applications where the interest lies in to perform predictions and/or analysis of covariance structures among time series. In this context, our interest is rather to assess the dynamic effects which are related to voxel activation. In order to do so, we develop three algorithms to simulate online-trajectories related to the state parameter and with those curves or simulated trajectories we compute a Monte Carlo evidence for voxel activation. Through two practical examples and two different types of assessments, we show that our method can be viewed for the practitioners as a reliable tool for fMRI data analysis. Despite all the examples and analysis are illustrated just for a single subject analysis, we also describe how more general group analysis can be implemented.

\end{abstract}

\noindent%
{\it Keywords:}  fMRI, Bayesian Analysis, Matrix-Variate Dynamic Linear Models, Monte Carlo Integration.

\spacingset{1.45}
\section{Introduction}
\label{sec:intro}

Magnetic resonance imaging (MRI) is a non-invasive technique that is used to create elaborate anatomical images of the human body. Specifically, this technique can be used to obtain detailed brain images that can help to identify different types of tissue such as, for example, white matter and gray matter, and can also be used to diagnose aneurysms and tumors. Another important facet of this technique is that it can be used to visualize dynamic or functional activity in the brain. Functional Magnetic Resonance Imaging (fMRI) can be described as a generalization of the MRI technique, where the focus is not just one high-resolution image of the brain, but rather a sequence
of low-resolution images that allows for identification, at least in an indirect way, of neuronal activity through the blood-oxygen-level dependent (BOLD) contrast (\cite{poldrack2011handbook}). The sequence of MRI images is obtained using an MRI scanner, essentially a huge magnet, which detects activity in the brain's different regions by tracing blood flows. Statistical models are very useful for analyzing the post-processed data that compose the sequence of images obtained in an fMRI experiment. The most popular statistical model used to identify a brain-region reaction to an external stimulus is the normal regression linear model, usually known as General Linear Model (GLM) in the fMRI literature, which assumes spatiotemporal independence among the data, an unrealistic assumption that can lead to incorrect inference. For instance, \cite{eklund2016cluster} evaluate the most common software packages (SPM, FSL and AFNI) for fMRI analysis, which have the GLM implemented as one of their analysis tools. Specifically, they evaluate that model for group analysis using real resting-state data and they find that, for a nominal family-wise error rate of 5\%, parametric statistical methods are shown to be conservative for voxel-wise inference and invalid for cluster-wise inference. However, there are other alternatives to model fMRI data, specially Bayesian models. Those alternatives usually account for the spatiotemporal structure present in this type of data. For example, \cite{zhang2016spatiotemporal} propose a spatiotemporal Nonparametric Bayesian approach to model both individual and group stages in one-step modeling. For posterior inference, they implement a Markov Chain Monte Carlo (MCMC) algorithm and additionally a suitable variational Bayes algorithm. \cite{eklund2017bayesian} propose a Bayesian heteroscedastic GLM with autoregressive linear noise and heteroscedastic noise innovations for single subject fMRI analysis. They develop an efficient MCMC algorithm that allows for variable selection among the regressors.  \cite{bezener2018bayesian} propose a Bayesian Spatiotemporal Modeling for single subject fMRI analysis. Their modeling includes a novel areal model to parcel voxels into clusters and also use an MCMC algorithm for posterior inference. \cite{yu2018bayesian} propose a Bayesian Variable Selection approach to model Complex-Valued fMRI data. They develop their models applying complex-valued spike-and-slab priors on the parameters associated with brain activation and perform posterior inference via MCMC algorithms. Those methods are just some of the most recent works published in the field. See \cite{zhang2015bayesian} and \cite{bowman2014brain} for a more complete review of statistical methods for fMRI data analysis.  \newline

In this work, we propose a local spatiotemporal Bayesian modeling for two-stage fMRI data analysis. However, here we do not resort to the traditional spatiotemporal models that are usually used to model fMRI data in the Bayesian contex. Instead, we use a Matrix-Variate Dynamic Linear Model (MDLM) (see \cite[p. 581]{west1997bayesian} for more details about these types of models), specifically, we employ the model proposed by \cite{quintana1985} and \cite{quintana1987multivariate} to model the BOLD response related to block and event-related experiments ( see \cite{kashou2014practical} for a detailed explanation of these kinds of designs). With this type of modeling, we can easily account for the temporal dependence through the latent component of the model which is most commonly known as the state equation. Despite the fact that, we take a voxel-wise approach, we model every voxel jointly with their nearest neighbors and thus the resulting covariance structure is used to perform the voxel-wise inference related to brain activation. This is what we call local spatiotemporal modeling. To define the clusters or neighborhoods related to every voxel, we just employ an Euclidean distance. Thus, the modeler has to define a fixed distance $r$ and all the voxels lying within that distance will be considered part of the cluster. Before explaining how this modeling is used to detect brain activation, it is worth mentioning that most of the work related to MDLM has been devoted to forecasting and analyzing covariance structure across several time series (\cite{quintana1987analysis}, \cite{aguilar2000bayesian}, \cite{fei2011bayesian}), however not much work has been developed regarding the inference of effect sizes related to regressors in an MDLM. \cite[p. 280]{west1997bayesian} briefly mention how to perform inference about the dynamic regression coefficient or state parameter, but only for a particular time $t$ within the interval of observed time.  In the case of an fMRI modeling, we would like to perform inference on the state parameter for an entire interval of time at once. In that sense, we propose three algorithms to draw on-line estimated trajectories of the state parameter and thus compute a measure of evidence via Monte Carlo integration. Hence, with that measure of evidence, we can perform inference about the state parameter or equivalently, in the context of fMRI modeling, about brain reaction. One algorithm is just an implementation of a forward-filtering-backward-sampling algorithm which was suggested by \cite{fruhwirth1994data} to draw state parameters for univariate dynamic linear models. In this matrix-variate setting, we use the same idea, but instead of estimating individual parameters for the covariance matrix of the state parameter in the Evolution equation (\ref{sec2:equ1}) using data augmentation, we use discount factors to deal with those quantities (\cite{Ameen1985normal}). The remain two algorithms can be described as forward-sampling schemes, in the sense that here we do not resort to the filtering distributions and just use the posterior distributions computed at each observed time $t$. 

Thus, at the individual stage, an MDLM is fitted for every subject in the sample, and analysis such as brain activation and/or contrast between tasks can be performed using any of the three algorithms mentioned above. At the group stage, the individual posterior distributions for the state parameters are combined in a suitable way, obtaining new distributions that represent the updated belief for the group effect. In this way, all of the analysis usually performed in a fMRI experiment (brain activation, task comparison, and group comparison) can be performed again by using any of the three algorithms proposed in this work. In order to show the performance of our method, we present two real data examples related to visual and motor task experiments respectively. We also assess our method in two ways: 1) following the same approach as in \cite{eklund2016cluster} using resting-state fMRI data from healthy controls, obtained from the 1000 Functional Connectomes Project \cite{biswal2010toward}. Here we define a fictitious stimulus and seek for the rate of false positives given the null hypothesis of no brain reaction. 2) Using simulated data obtained from the R package \textbf{neuRosim} \citep{welvaert2011neurosim}. \\
In the next section, we give a brief description of the Matrix-Variate Dynamic Linear Model and its posterior distribution under a matrix Normal/inverse Wishart prior distribution. We also describe the modeling of the BOLD response using this type model for single and group analysis. Section \ref{sec:3} describes the three algorithms and computation of Monte Carlo evidence for voxel activation. In section \ref{sec:4} two examples for block and event-related designs are presented to illustrate the performance of our proposal. In section \ref{sec:5} an assessment is performed to evaluate the capacity of our method to deal with false-positive activations. In the final section, we present some concluding remarks and potential open problems for future development.

\section{MVDLM and fRMI Modeling}
\label{sec:meth}

The general theory of the MVDLM is presented in \cite{quintana1985}, \cite{quintana1987multivariate} and \cite{west1997bayesian}. Despite this model having been conceived for forecasting, in this work it is used for a different purpose: model fMRI data related to block and event-related designs through a regression structure in order to identify brain reaction. In other words, we focus our interest on the estimation of, and inference on, the state parameter. Thus, suppose that we have a $q\times 1$ vector $\mathbf{Y}_t$, which can be modeled in terms of observation and evolution or state equations as follows.

\begin{equation} \label{sec2:equ1}
\begin{array}{lccl}
\text{Observation:}\ &\mathbf{Y}_t & =& \mathbf{F}^{'}_t\mathbf{\Theta}_t + \bm{\nu}^{'}_t \\
 \text{Evolution:}\ &\mathbf{\Theta}_t& =& \mathbf{G}_t\mathbf{\Theta}_{t-1} + \mathbf{\Omega}_t.\\
\end{array}
\end{equation}

Where, for each $t$ we have a $q \times 1$ vector $\bm{\nu}^{'}_t$ of observational errors, a $p\times q$ matrix $\mathbf{\Theta}_t$ of state parameters, a $p\times q$ matrix $\mathbf{\Omega}_t$ of evolution errors. The $1 \times p$  and $p \times p$ matrices $\mathbf{F}^{'}_t$ and $\mathbf{G}_t$ respectivelly are common to each of the $q$ univariate DLMs. The covariates related to the design being used, either a block or an event-related design as well as other characteristics of the subjects, can be included in the columns of $\mathbf{F}^{'}_t$. It is supposed that $\nu_t \sim N_q\left[\mathbf{0}, V_t \mathbf{\Sigma}\right]$,  independently over time, where $\mathbf{\Sigma}$ defines the cross-sectional covariance structure for the multivariate model and $V_t$ is a known observational scale factor. For the random matrix $\Omega_t$ the distribution is given by $\Omega_t \sim N_{pq}\left[\mathbf{0}, \mathbf{W}_t, \mathbf{\Sigma}\right]$, a matrix-variate normal distribution with mean matrix $\mathbf{0}$, left $p\times p$ variance matrix $\mathbf{W}_t$ and right variance matrix $\mathbf{\Sigma}$ (\cite{dawid1981some}). It is also supposed the initial prior for $\mathbf{\Theta}_0$ and $\mathbf{\Sigma}$ is matrix normal/inverse Wishart, namely $(\mathbf{\Theta}_0, \mathbf{\Sigma}|D_0) \sim NW_{n_0}^{-1}\left[\mathbf{m}_0,\mathbf{C}_0, \mathbf{S}_0 \right],$ for some known defining parameters $\mathbf{m}_0$, $\mathbf{C}_0$, $\mathbf{S}_0$ and $n_0$ (\citep{quintana1987multivariate}[Chapter 3]). $D_t = \{\mathbf{Y}_0, \mathbf{Y}_1, \ldots, \mathbf{Y}_t\}$ are the data observed at time $t$. 

\subsection{Voxel-wise Individual Analysis}

One of the main objectives in an fMRI experiment is to identify a brain reaction in response to some controlled external stimuli, in other words, to look for changes in the BOLD signal related to some experimental manipulation. In order to do that, one can use a block design, an event-related design or even possible, a new design from the composition of those two (\cite{kashou2014practical}). The modeling we propose here relies on something called the expected BOLD response, which is obtained as the convolution of the stimulus time series $f$, which is obtained from the design being used, and the hemodynamic response function $h$ (HRF), which can be interpreted as the picture over time of the change of the BOLD signal. In the examples and analysis presented here we use the canonical HRF, which is defined as the difference (between) of two gamma densities. However, is it worth mentioning that any other form from the available options for the HRF can be used. This subject is discussed in depth in \cite{poldrack2011handbook} and \cite{penny2011statistical}. In general, the expected BOLD response can be obtained as follows:

\begin{equation}\label{equ2}
x(t)=(h\ast f)(t)=\int h(\tau)f(t-\tau)d\tau.
\end{equation}

Thus, the aim is to identify the fMRI time series that matches the expected BOLD response. For that purpose, we model the observed BOLD response as a linear function of the expected BOLD response using the model (\ref{sec2:equ1}). 

\subsubsection{Individual Modeling}

Let $y_{\scaleto{[i,j,k],t,1\mathstrut}{6pt}}^{*(z)}$ and $x^{(z)}(t)$ be the observed and expected BOLD response, respectively at brain position $\{i,j,k\}$, time $t$ and subject $z$, for $i=1, \ldots, d_1$, $j=1,\ldots,d_2$, $k=1,\ldots,d_3$, $t=1,\ldots, T$ and $z=1\ldots,N_g$. Something that can be noticed from this last definition is that $x^{(z)}(t)$ is supposed to be the same for all the locations in the brain image, which means, it is supposing that the BOLD response is the same in all brain regions. This may be an unrealistic assumption, but it is one that works well in practice. Let the vector

\begin{equation}\label{sec2:equ2}
 \mathbf{Y}_{\scaleto{[i,j,k]t\mathstrut}{6pt}}^{(z)}=\left( \begin{array}{c}
y_{\scaleto{[i,j,k],t,1\mathstrut}{6pt}}^{*(z)}\\ 
y_{\scaleto{[i+1,j,k],t,2\mathstrut}{6pt}}^{(z)}\\ 
y_{\scaleto{[i-1,j,k],t,3\mathstrut}{6pt}}^{(z)}\\
y_{\scaleto{[i,j+1,k],t,4\mathstrut}{6pt}}^{(z)}\\
y_{\scaleto{[i,j-1,k],t,5\mathstrut}{6pt}}^{(z)}\\ 
y_{\scaleto{[i,j,k+1],t,6\mathstrut}{6pt}}^{(z)}\\ 
y_{\scaleto{[i,j,k-1],t,7\mathstrut}{6pt}}^{(z)}
\end{array}\right)_{\scaleto{1\times q\mathstrut}{5pt}}
\end{equation}represent the cluster or neighborhood of size $q=7$ of the voxel at position $\{i,j,k\}$ and time $t$. Thus, we model $\mathbf{Y}_{\scaleto{[i,j,k]t\mathstrut}{6pt}}^{(z)}$ using the model (\ref{sec2:equ1}), with $\mathbf{G}_t=\mathbf{I}_p$ and $\mathbf{F}_t^{'(z)}=(x^{(z)}_1(t),\ldots,x^{(z)}_{p}(t))$, where $x^{(z)}_1(t),\ldots, x^{(z)}_{p}(t)$ are related to different types of stimuli or tasks performed in the fMRI experiment. What we intend with this type of modeling is to capture the covariance structure within the cluster of voxels through the matrix $\mathbf{\Sigma}^{(z)}$. The criterion to define the cluster form is based on the Euclidean distance, where the distance between the voxel $V^{*}$ and the neighboring voxel $V_{\sim}$ is given by $d(V^{*}, V_{\sim})\leq r$. In figure \ref{cap3:fig0} (right panel), we can see a graphical illustration of a neighborhood of voxels related to the observed BOLD response vector (\ref{sec2:equ2}) and its corresponded matrix parameter on equation (\ref{sec2:equ3}), where $r=1$ and consequently $q=7$. In the R-package\footnote{\url{https://github.com/JohnatanLAB/BayesDLMfMRI}} where this method has been implemented, a range of different $r$ values can be set by the users: $r=1, 2, 3, 4$, which consequently leads to $q=7, 19, 27, 33$.

\begin{table}[H]
\begin{figure}[H]
  \centering
\begin{center}
\begin{tabular}{ccc}
\includegraphics[width=.25\textwidth]{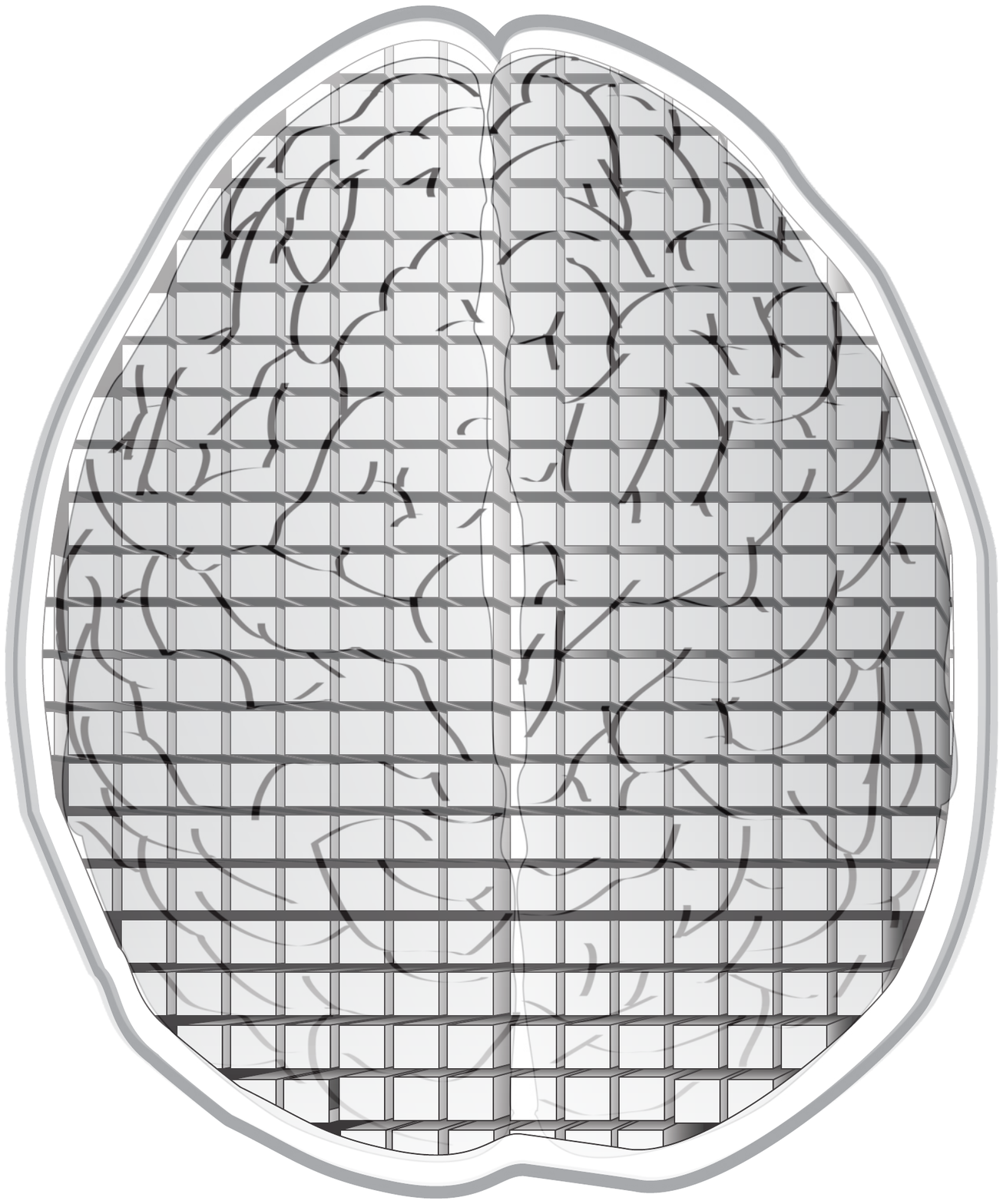}&\includegraphics[width=.25\textwidth]{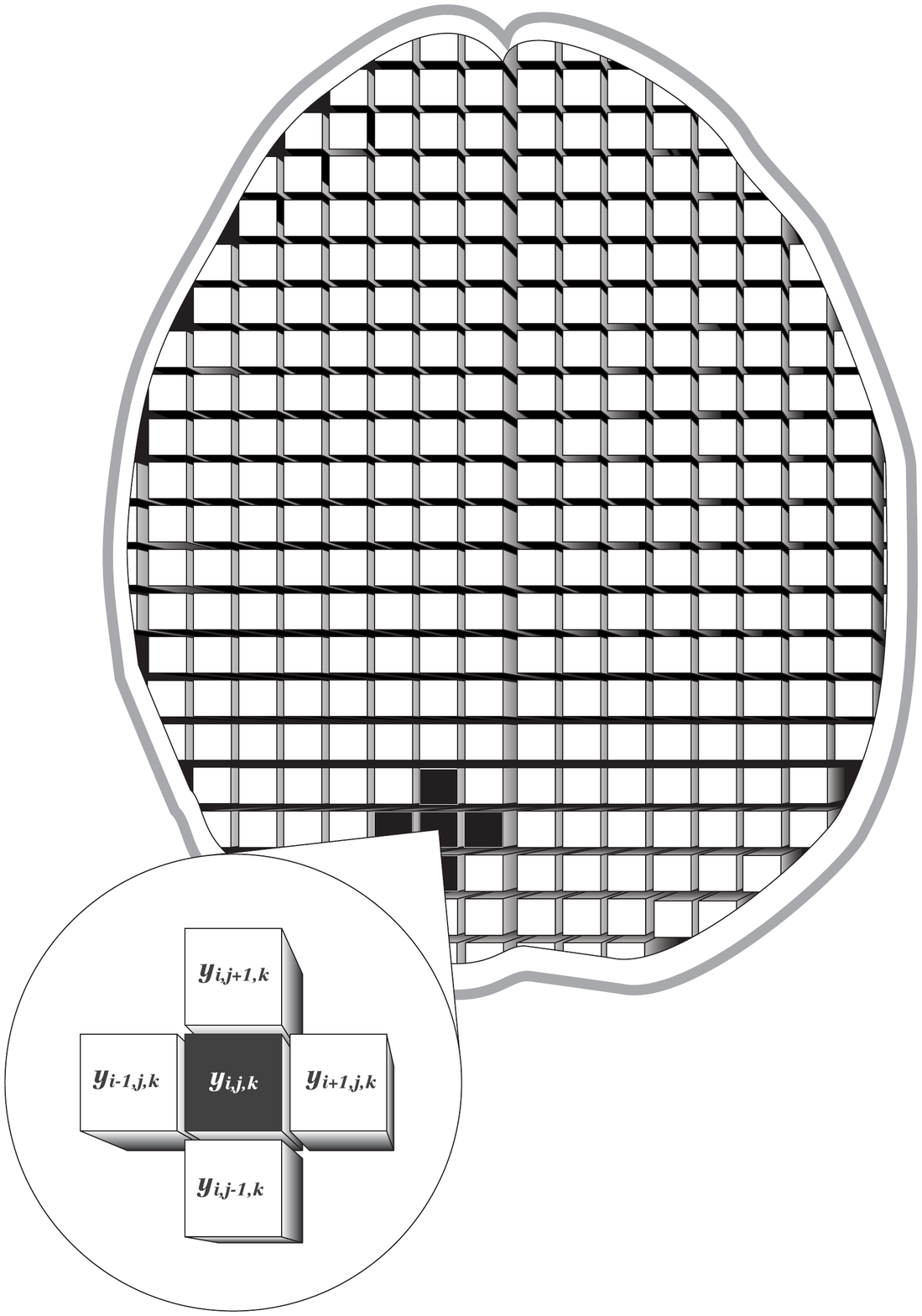}&\includegraphics[width=.25\textwidth]{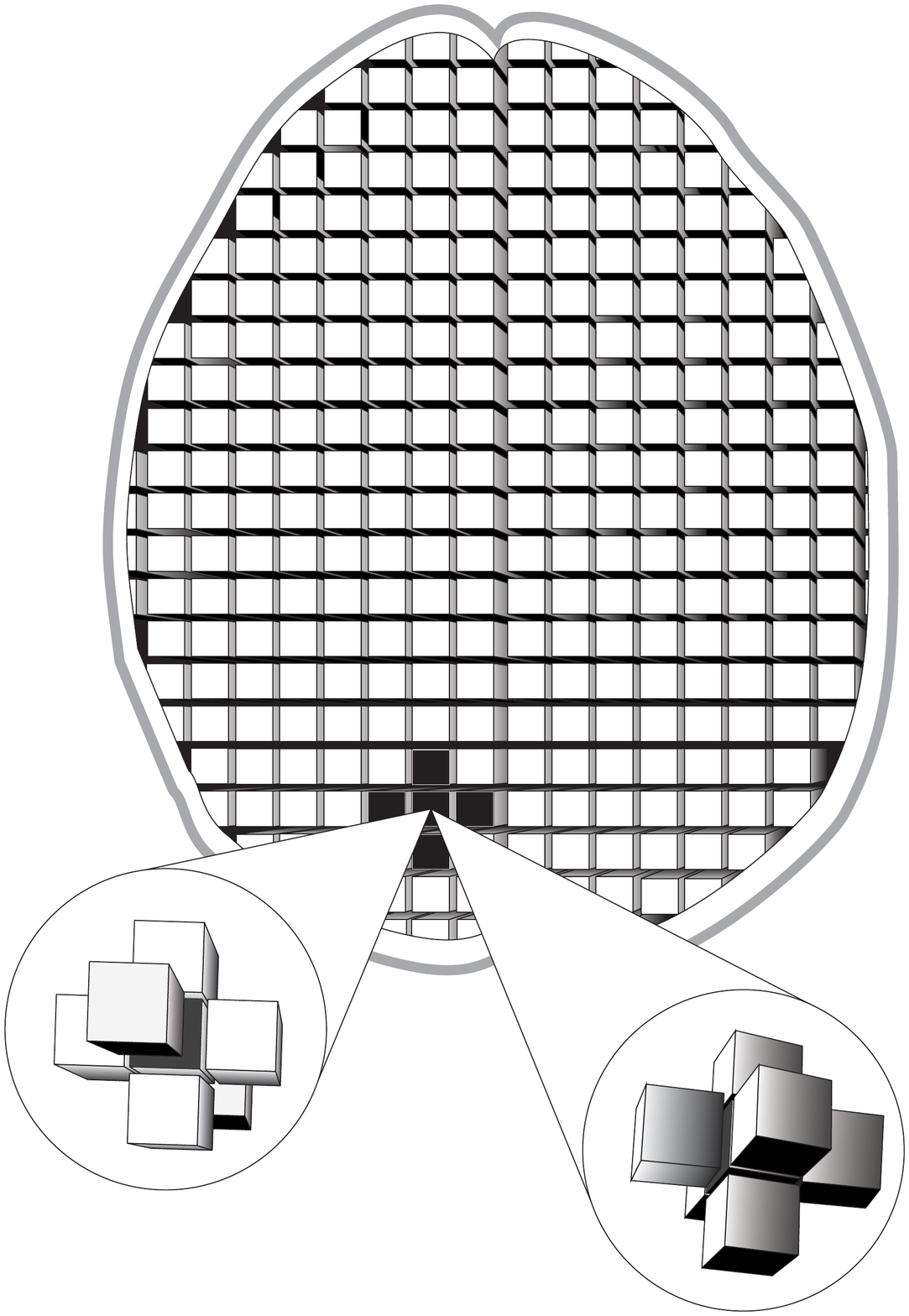}\\
\end{tabular}
\end{center}
  \caption{Graphical illustration of a neighborhood of voxels, for $r=1$.}
  \label{cap3:fig0} 
\end{figure}
\end{table} 

\begin{equation}\label{sec2:equ3}
 \mathbf{\Theta}_{\scaleto{[i,j,k]t\mathstrut}{6pt}}^{(z)}=\left( \begin{array}{ccccccc}
\theta_{\scaleto{[i,j,k],t,1,1\mathstrut}{6pt}}^{*(z)}&\theta_{\scaleto{[i,j,k],t,1,2\mathstrut}{6pt}}^{(z)}&\theta_{\scaleto{[i,j,k],t,1,3\mathstrut}{6pt}}^{(z)}&\theta_{\scaleto{[i,j,k],t,1,4\mathstrut}{6pt}}^{(z)}&\theta_{\scaleto{[i,j,k],t,1,5\mathstrut}{6pt}}^{(z)}&\theta_{\scaleto{[i,j,k],t,1,6\mathstrut}{6pt}}^{(z)}&\theta_{\scaleto{[i,j,k],t,1,7\mathstrut}{6pt}}^{(z)}\\ 
\vdots & \vdots & \vdots & \vdots & \vdots & \vdots & \vdots\\
\theta_{\scaleto{[i,j,k],t,p,1\mathstrut}{6pt}}^{*(z)}&\theta_{\scaleto{[i,j,k],t,p,2\mathstrut}{6pt}}^{(z)}&\theta_{\scaleto{[i,j,k],t,p,3\mathstrut}{6pt}}^{(z)}&\theta_{\scaleto{[i,j,k],t,p,4\mathstrut}{6pt}}^{(z)}&\theta_{\scaleto{[i,j,k],t,p,5\mathstrut}{6pt}}^{(z)}&\theta_{\scaleto{[i,j,k],t,p,6\mathstrut}{6pt}}^{(z)}&\theta_{\scaleto{[i,j,k],t,p,7\mathstrut}{6pt}}^{(z)}
\end{array}\right).
\end{equation}

From here to the rest of the article, we change the notation related to the location subscript $[i,j,k]$ by $v$.

\subsubsection*{Posterior inference}

Under the assumptions and prior distributions defined above and following the results in \cite{dawid1981some}, it can be shown that the joint posterior distribution of the state parameter $\mathbf{\Theta}_{vt}^{(z)}$ and the covariance matrix $\mathbf{\Sigma}_v^{(z)}$ is given by 

\begin{equation}\label{sec2:equ4}
(\mathbf{\Theta}_{vt}^{(z)}, \mathbf{\Sigma}_v^{(z)}|D_{vt})\sim NW^{-1}_{n_t}[\mathbf{m}_{vt}^{(z)}, \mathbf{C}_{vt}^{(z)}, \mathbf{S}_{vt}^{(z)}],
\end{equation}with

\[ \mathbf{m}_{vt}^{(z)}= \mathbf{m}_{v,t-1}^{(z)} + \mathbf{A}_{vt}^{(z)}\mathbf{e}_{vt}^{'(z)} \ \ \textbf{and} \ \ \mathbf{C}_{vt}^{(z)}= \mathbf{C}_{v,t-1}^{(z)} + \mathbf{W}_{vt}^{(z)} - \mathbf{A}_{vt}^{(z)}\mathbf{A}_{vt}^{'(z)}Q_{vt}^{(z)} , \]
\[n_t=n_t + 1 \ \ \textbf{and} \ \ \mathbf{S}_{vt}^{(z)} = n_t^{-1}[n_{t-1}\mathbf{S}_{v,t-1}^{(z)} + \mathbf{e}^{'(z)}_{vt}\mathbf{e}_{vt}^{(z)}/Q_{vt}^{(z)}]\]where

\[ Q_{vt}^{(z)} = V_t + \mathbf{F}_t^{'(z)}[\mathbf{C}_{v,t-1}^{(z)} + \mathbf{W}_{vt}^{(z)} ]\mathbf{F}_t^{(z)}, \ \ \mathbf{A}_{vt}^{(z)} = [\mathbf{C}_{v,t-1}^{(z)} + \mathbf{W}_{vt}^{(z)} ]F_t^{(z)}/Q_{vt}^{(z)}, \ \ \textbf{and} \ \ \mathbf{e}_{vt}^{(z)} = \mathbf{Y}_{vt}^{(z)}-\mathbf{f}_{vt}^{(z)}.\]

$\mathbf{f}_{vt}^{(z)}$ is the mean of the predictive distribution $p(\mathbf{Y}_{vt}^{(z)}|D_{v,t-1}^{(z)})$. Hence, the marginal posterior distribution for the state parameter is given by

\begin{equation}\label{sec2:equ5}
(\mathbf{\Theta}_{vt}^{(z)}|D_{vt}^{(z)})\sim T_{n_t}[\mathbf{m}_{vt}^{(z)}, \mathbf{C}_{vt}^{(z)}, \mathbf{S}_{tv}^{(z)}],
\end{equation}

which is known as a matrix $T$ distribution with $p\times q$ mean matrix $\mathbf{m}_{vt}^{(z)}$, $p\times p$ left variance matrix $\mathbf{C}_{vt}^{(z)}$, and $q\times q$ right variance matrix $\mathbf{S}_{vt}^{(z)}$. To deal with the unknown matrix parameter $\mathbf{W}_{vt}^{(z)}$, we adopt the discount model approach proposed by \cite{Ameen1985normal} in the same manner as in \cite{west1985dynamic}. Thus, we express the system variace matrix as $\mathbf{W}_{vt}^{(z)} = \mathbf{B_t}\mathbf{C}_{v,t-1}^{(z)}\mathbf{B_t} - \mathbf{C}_{v,t-1}^{(z)}$, where $\mathbf{B_t}$ is an $p \times p$ diagonal matrix of positive discount factors $1/\beta_{it}^{1/2}$, for $i=1,\ldots,p$ and $0<\beta_{it}\leq 1$. As we show in one of the examples below, those discount factors are a key component of this modeling in the sense that they can affect the sensibility of detection of brain activation. \newline A reasonable approximation for the posterior distribution (\ref{sec2:equ5}) when $n_t\geq 30$ is given by

\begin{equation}\label{sec2:equ6}
(\mathbf{\Theta}_{vt}^{(z)}|D_{vt}^{(z)}) \widesim{approx} N_{pq}[\mathbf{m}_{vt}^{(z)}, \mathbf{C}_{vt}^{(z)}, \mathbf{S}_{vt}^{(z)}].
\end{equation}

We have two main considerations to justify working only with posterior distributions for $n_t\geq 30$. The first is that as we use vague prior distributions at $t=0$, the sequential update process takes some period of time before reaching a posterior distribution dominated by the data. Then, the first posterior distributions (i.e., for $t<30$) could be considered irrelevant for the analysis. The second consideration is simply that dealing with normal distributions simplifies the mathematical work when dealing with linear combinations, which are quite common in this work when  inference is performed on the matrix parameter $\mathbf{\Theta}_{vt}^{(z)}$ to detect brain activation. \\
As mentioned above, a brain activation related to the stimulus or task $l$ (for $l=1,\ldots,p$) is identified when the expected BOLD response ($x^{(z)}_l(t)$) matches the observed BOLD response ($ \mathbf{Y}_{\scaleto{vt\mathstrut}{6pt}}^{(z)}$). In other words, a brain activation in the cluster of voxels $v$ related to the $l$-th task or stimulus is equivalent to the $l$-th row of $\mathbf{\Theta}_{vt}^{(z)}$ being positive. In this sense, we define three different ways to evaluate brain activation, which are related to the following variables obtained from the components of $\mathbf{\Theta}_{vt}^{(z)}$

\begin{equation}\label{sec2:equ7}
\text{Marginal effect:} \ \ \theta^{*(z)}_{vt,l} = \theta_{\scaleto{v,t,l,1\mathstrut}{6pt}}^{*(z)},
\end{equation}

\begin{equation}\label{sec2:equ8}
\text{Average cluster effect:} \ \ \bar{\theta}_{vt,l}^{(z)} = \frac{1}{q} \left[ \theta_{\scaleto{v,t,l,1\mathstrut}{6pt}}^{*(z)} +  \sum \limits_{n=2}^{q}  \theta_{\scaleto{v,t,l,n \mathstrut}{6pt}}^{(z)}\right],
\end{equation}

\begin{equation} \label{sec2:equ9}
\text{Joint effect:} \ \ \boldsymbol{\theta}_{vt,l}^{(z)}=\left(\theta_{\scaleto{v,t,l,1\mathstrut}{6pt}}^{*(z)}, \theta_{\scaleto{v,t,l,2\mathstrut}{6pt}}^{(z)}, \ldots,\theta_{\scaleto{v,t,l,q\mathstrut}{6pt}}^{(z)}\right).
\end{equation}

The marginal effect (\ref{sec2:equ7}) is when just a marginal distribution from (\ref{sec2:equ6}) is considered, in other words when the correlation captured by $\mathbf{S}_{vt}^{(z)}$ is ignored. It can be shown that under particular considerations a DLM is equivalent to a traditional static regression linear model (or GLM in the fMRI context), so an analysis using marginal effects can yield results pretty similar to those obtained when performing a voxel-wise approach using a GLM. Nevertheless, this depends on the variance value of the state parameter ($\mathbf{W}_{vt}^{(z)}=0$). The average (\ref{sec2:equ8})  and joint (\ref{sec2:equ9}) effects are more interesting in the sense that they take advantage of the information contained within the entire cluster of voxels.

From the properties of the matrix normal distribution, the distributions of (\ref{sec2:equ7}), (\ref{sec2:equ8}) and (\ref{sec2:equ9}) are given by

\begin{equation}\label{sec2:equ10}
\theta_{vt,l}^{*(z)}|D_{\scaleto{vt\mathstrut}{6pt}}^{*(z)}\sim N(m_{\scaleto{vt,l,1\mathstrut}{6pt}}^{*(z)}, C_{\scaleto{vt,l,l\mathstrut}{6pt}}^{(z)} S_{\scaleto{vt,1,1\mathstrut}{6pt}}^{(z)}),
\end{equation}

\begin{equation}\label{sec2:equ11}
\bar{\theta}_{\scaleto{vt,l\mathstrut}{6pt}}^{(z)}|D_{\scaleto{vt\mathstrut}{6pt}}^{(z)}\sim N(\bar{m}_{\scaleto{vt,l\mathstrut}{6pt}}^{(z)}, \bar{S}_{\scaleto{vt,l\mathstrut}{6pt}}^{(z)}),
\end{equation}

\begin{equation}\label{sec2:equ12}
\boldsymbol{\theta}_{\scaleto{vt,l\mathstrut}{6pt}}^{(z)}|D_{\scaleto{vt\mathstrut}{6pt}}^{(z)}\sim N_q(\boldsymbol{m}_{\scaleto{vt,l\mathstrut}{6pt}}^{(z)}, C_{\scaleto{vt,l,l\mathstrut}{6pt}}^{(z)}\boldsymbol{S}_{\scaleto{vt\mathstrut}{6pt}}^{(z)}),
\end{equation}

where $C_{\scaleto{vt,l,l\mathstrut}{6pt}}^{(z)}$ and $S_{\scaleto{vt,n,n\mathstrut}{6pt}}^{(z)}$ are the elements on the main diagonal of the matrices $\boldsymbol{C}_{vt}^{(z)}$ and $\boldsymbol{S}_{\scaleto{vt\mathstrut}{6pt}}^{(z)}$ respectively, $\bar{m}_{\scaleto{vt,l\mathstrut}{6pt}}=\frac{1}{q}\sum \limits_{n=1}^{q}m_{\scaleto{vt,l,n\mathstrut}{6pt}}$, $\bar{S}_{\scaleto{vt,l\mathstrut}{6pt}}=\frac{1}{q^2}\left[\sum \limits_{n=1}^{q}C_{vt,l,l}S_{vt,n,n} + \sum \limits_{n\neq n^{'}} C_{vt,l,l}S_{vt,n,n^{'}} \right]$ and $D_{\scaleto{vt\mathstrut}{6pt}}^{*(z)}$ is the data related to first component of the vector (\ref{sec2:equ2}). With the distributions (\ref{sec2:equ10}), (\ref{sec2:equ11}) and (\ref{sec2:equ12}), at least from our knowledge, it is only possible to perform inference (about (\ref{sec2:equ7}), (\ref{sec2:equ8}) and (\ref{sec2:equ9})) for a particular fixed time $t$, which in the context of fMRI means inference related to only one brain image. However, in fMRI data analysis, it is more relevant to perform an inference procedure at once for the entire group of brain images just to understand brain dynamics related to the controlled stimulation presented on the fMRI experiment. In the next section, we present three algorithms that allow us to draw dynamic trajectories of the state parameter related to any of (\ref{sec2:equ7}), (\ref{sec2:equ8}) and (\ref{sec2:equ9}) and hence using those trajectories to compute a measure of evidence for voxel activation on the entire observed interval of brain images. But before we get there, we first offer a brief explanation on how to get versions of (\ref{sec2:equ7}), (\ref{sec2:equ8}) and (\ref{sec2:equ9}) for the group stage analysis.

\subsection{Voxel-wise Group Analysis}

Now, we describe the fMRI group analysis where the focus is to detect an average group activation (single-group analysis) and/or compare voxel activation between two groups (e.g. patients vs. controls). Here we take any of the posterior distributions (\ref{sec2:equ10}), (\ref{sec2:equ11}), or (\ref{sec2:equ12}), depending on the case, as an input for this stage. For instance, let's suppose that a sample of $N_g$ subjects are part of an fMRI experiment where $p$ stimuli are presented. Thus, taking advantage of the properties of normal distriution, we obtain the following distributions for the group stage:

\begin{equation}\label{sec2:equ13}
\text{Marginal group effect:} \ \ \bar{\theta}_{\scaleto{vt,l\mathstrut}{6pt}}^{*(g)} \sim N(\bar{m}_{\scaleto{vt,l\mathstrut}{6pt}}^{*(g)}, \bar{S}_{\scaleto{vt,l\mathstrut}{6pt}}^{(g)}),
\end{equation}

\begin{equation}\label{sec2:equ14}
\text{Average cluster group effect:} \ \ \bar{\bar{\theta}}_{\scaleto{vt,l\mathstrut}{6pt}}^{(g)}\sim N(\bar{\bar{m}}_{\scaleto{vt,l\mathstrut}{6pt}}^{(g)}, \bar{\bar{S}}_{\scaleto{t,l\mathstrut}{6pt}}^{(g)}),
\end{equation}

\begin{equation}\label{sec2:equ15}
\text{Joint group effect:} \ \ \boldsymbol{\bar{\theta}}_{\scaleto{vt,l\mathstrut}{6pt}}^{(g)}\sim N_q(\boldsymbol{\bar{m}}_{\scaleto{vt,l\mathstrut}{6pt}}^{(g)}, \boldsymbol{\bar{S}}_{\scaleto{t\mathstrut}{6pt}}^{(g)}),
\end{equation}

where,

\begin{equation*}
\begin{array}{ll}
\bar{m}_{\scaleto{vt,l\mathstrut}{6pt}}^{*(g)}= \frac{1}{N_g}\sum \limits_{z=1}^{N_g}m_{\scaleto{vt,l,1\mathstrut}{6pt}}^{*(z)}& \bar{S}_{\scaleto{vt,l\mathstrut}{6pt}}^{(g)}=\frac{1}{N_{g}^2}\sum \limits_{z=1}^{n_g}C_{\scaleto{vt,l,l\mathstrut}{6pt}}^{(z)}*S_{\scaleto{vt,1,1\mathstrut}{6pt}}^{(z)},\\
\\
\bar{\bar{m}}_{\scaleto{vt,l\mathstrut}{6pt}}^{(g)}=\frac{1}{N_{g}}\sum \limits_{z=1}^{n_g}\bar{m}_{\scaleto{vt,l\mathstrut}{6pt}}^{(z)}& \bar{\bar{S}}_{\scaleto{vt,l\mathstrut}{6pt}}^{(g)}=\frac{1}{N_{g}^2}\sum \limits_{z=1}^{N_g}\bar{S}_{\scaleto{vt,l\mathstrut}{6pt}}^{(z)},\\
\\
\boldsymbol{\bar{m}}_{\scaleto{vt,l\mathstrut}{6pt}}^{(g)}=\frac{1}{N_{g}}\sum \limits_{z=1}^{N_g}\boldsymbol{m}_{\scaleto{vt,l\mathstrut}{6pt}}^{(z)}& \boldsymbol{\bar{S}}_{\scaleto{vt\mathstrut}{6pt}}^{(g)}=\frac{1}{N_{g}^2}\sum \limits_{z=1}^{n_g}C_{\scaleto{vt,l,l\mathstrut}{6pt}}^{(z)}\boldsymbol{S}_{\scaleto{vt\mathstrut}{6pt}}^{(z)}.
\end{array}
\end{equation*} 

Let's suppose now there are two groups we want to compare. For the patiens' group (group $\boldsymbol{A}$), we have $N_A$ individuals and for the controls' group (group $\boldsymbol{B}$), we have $N_B$ individuals. Thus, for instance, if we decide to compare the joint effect between the two groups, the most obvious way to do so is by computing the distribution for the variable $\boldsymbol{\bar{\theta}}_{\scaleto{vt,l\mathstrut}{6pt}}^{(AB)}= \boldsymbol{\bar{\theta}}_{\scaleto{vt,l\mathstrut}{6pt}}^{(A)}-\boldsymbol{\bar{\theta}}_{\scaleto{vt,l\mathstrut}{6pt}}^{(B)}$, which in this case is given by $N_q(\boldsymbol{\bar{m}}_{\scaleto{vt,l\mathstrut}{6pt}}^{(A)}-\boldsymbol{\bar{m}}_{\scaleto{vt,l\mathstrut}{6pt}}^{(B)}, \boldsymbol{\bar{S}}_{\scaleto{t\mathstrut}{6pt}}^{(A)}+\boldsymbol{\bar{S}}_{\scaleto{t\mathstrut}{6pt}}^{(B)})$. The analysis is analogous for the marginal and average effects. As it was mentioned above, under this setting, it is only possible to perform any type of inference for a particular fixed time $t$. In the next section, we present three algorithms that help us to overcome this limitation.

\section{Algorithms}
\label{sec:3}

In this section we present three different algorithms to draw on-line trajectories of the estate parameter. The first two algorithms only depends on each posterior distribution obtained at each time $t$, whereas the third one is based on the filtered distribution $p(\mathbf{\Theta}_{v,t-j}^{(z)}, \mathbf{\Sigma}^{(z)}|D_{vt}^{(z)})$, for $1<j \leq t$. For the examples and  assessment presented on sections (\ref{sec:4}) and (\ref{sec:5}), we implement model (\ref{sec2:equ1}) along these algorithms in an R package\footnote{\url{https://github.com/JohnatanLAB/BayesDLMfMRI}} (\cite{Rcite}) with the help of the RcppArmadillo (\cite{RcppArma}) and parallel packages to speed up the computation time. We also make use of the oro.nifti \citep{oro.nifti} and neurobase \citep{neurobase} packages to import and export the fMRI data on nii.gz format and plot the evidence activation maps. The FLS software (\cite{jenkinson2012fsl}) is used when preprocessing of the fMRI data is needed.

\subsection{ Forward Estimated Trajectories Sampler or FEST algoritm}

Let's suppose we are interested in detect brain activation using the distribution (\ref{sec2:equ11}). Then, we define the online estimated trajectory of $\boldsymbol{\theta}_{\scaleto{vt,l\mathstrut}{6pt}}^{(z)}$ as the vector:
 \[\gamma_{v,l}^{(z)} = \{ E(\boldsymbol{\theta}_{\scaleto{vt_1,l\mathstrut}{6pt}}^{(z)}|D_{\scaleto{vt_1\mathstrut}{6pt}}^{(z)}), E(\boldsymbol{\theta}_{\scaleto{vt_2,l\mathstrut}{6pt}}^{(z)}|D_{\scaleto{vt_2\mathstrut}{6pt}}^{(z)}), \ldots, E(\boldsymbol{\theta}_{\scaleto{vT,l\mathstrut}{6pt}}^{(z)}|D_{\scaleto{vT\mathstrut}{6pt}}^{(z)})\},\] for $t_1 \geq 30$.
 In case of a significant match between the expected and observed BOLD responses, one would expect every component of the vector $\gamma_{v,l}^{(z)}$ being positive.
  In this sense, we must compute a measure of evidence for brain activation at voxel $v$, such as $p(\gamma_{v,l}^{(z)}>0)$. For instance, if $p(\gamma_{v,l}^{(z)}>0)>\alpha$, where $\alpha$ is a threshold defined by the user (e.g. $\alpha=0.95$), then one would conclude that there is significant evidence of a match between the expected and the observed BOLD responses. In other words, there is a brain activation related to the $l$-th stimulus or task in the cluster related to the voxel $v$. The functional variable $\gamma_{v,l}^{(z)}$ can be defined in the same way in case of using either (\ref{sec2:equ10}) or (\ref{sec2:equ12}).
Thus, taking advantage of the sequential update process related to the posterior distribution (\ref{sec2:equ6}), we build an algorithm to draw $\gamma_{v,l}^{(z)}$ variables. We name it as Forward Estimated Trajectories Sampler (FETS).  

\begin{algorithm}
\caption{FETS}\label{euclid}
\begin{algorithmic}[1]
\Procedure{ $ \textsf{for}\ \ k=1\ldots N$, $p(\theta^{(z)}_{vt,l}|D_{vt}^{(z)})$ being any of (\ref{sec2:equ10}), (\ref{sec2:equ11}) or (\ref{sec2:equ12}) and $\boldsymbol{s}_{\scaleto{vt\mathstrut}{6pt}}^{(z)}$ a posterior estimation related to $\mathbf{\Sigma}_{vt}$}{}
\State Draw $\theta^{(z,k)}_{vt,l}$ from $p(\theta^{(z)}_{vt,l}|D_{vt}^{(z)})$ for $t=1,\ldots,T$ and $l=1, \ldots, p$
\State Draw $\nu_{vt}^{(z,k)}$ from $N_q[0,\boldsymbol{s}_{\scaleto{vt\mathstrut}{6pt}}^{(z)}]$ for $t=1,\ldots,T$
\State Compute $\tilde{Y}^{(z,k)}_{vt}= \sum \limits_{l=1}^{p}\theta^{(z,k)}_{vt,l}x^{(z)}_l(t) +\nu_{vt}^{(z,k)}$ for $t=1,\ldots,T$
\State Compute $p(\theta^{(z,k)}_{vt,l}|\tilde{D}_{vt}^{(z)})$ for $t=1,\ldots,T$, where $\tilde{D}_{vt}^{(z)} = \{\tilde{Y}^{(z,k)}_{v1}, \ldots, \tilde{Y}^{(z,k)}_{vt}\}$
\State Let $\gamma_{v,l}^{(z,k)}=\{E(\theta^{(z,k)}_{v1,l}|\tilde{D}_{v1}^{(z)}), \ldots, E(\theta^{(z,k)}_{vT,l}|\tilde{D}_{vT}^{(z)})\}$ for $l=1, \ldots, p$
\EndProcedure
\end{algorithmic}
\end{algorithm}

Then, a measure of evidence for the activation of voxel $v$ is computed as

\begin{equation}\label{moteCarlo1}
p(\gamma_{v,l}^{(z)}>0) = E(\mathds{1}_{(\gamma_{v,l}^{(z)}>0)})\approx \frac{\sum\limits_{k=1}^{N} \mathds{1}_{(\gamma_{v,l}^{(z,k)}>0)}}{N},
\end{equation}

where $\mathds{1}_{(A)}$ is the indicator function related to the event $A$. The estimation $\boldsymbol{s}_{\scaleto{vt\mathstrut}{6pt}}^{(z)}$ depends on the chioce for $p(\theta^{(z)}_{vt,l}|D_{vt}^{(z)})$. Thus, for any of (\ref{sec2:equ10}), (\ref{sec2:equ11}) or (\ref{sec2:equ12}),  $\boldsymbol{s}_{\scaleto{vt\mathstrut}{6pt}}^{(z)}$ can be replaced by $S_{\scaleto{vt,1,1\mathstrut}{6pt}}^{(z)}$,  $\frac{1}{q^2}\left[\sum \limits_{n=1}^{q}S_{vt,n,n}^{(z)} + \sum \limits_{n\neq n^{'}}S_{vt,n,n^{'}}^{(z)} \right]$ or $\mathbf{S}_{vt}^{(z)}$ respectively. It is also possible to compute a measure of evidence for a contrast between two different tasks or stimuli. For instance, let's suppose one is interested in the comparison $\gamma_{v,l}^{(z)}>\gamma_{v,l^{'}}^{(z)}$. In other words, he or she wants to investigate whether the brain activation related to the task $l$ is stronger than that related to task $l^{'}$. To answer that question, one can compute $p(\gamma_{v,l}^{(z)} - \gamma_{v,l^{'}}^{(z)}>0)$ taking the algorithm's output in the step 6 and using Mote Carlo integration analogously as in (\ref{moteCarlo1}). Besides of the use of the \textbf{FEST} algorithm to compute measures of evidence for brain activation, it also generates other useful outputs such as $\{\tilde{Y}^{(z,k)}_{v1},\ldots, \tilde{Y}^{(z,k)}_{vt}\}$ and $\{\mathbf{S}_{v1}^{(z,k)}, \ldots, \mathbf{S}_{vt}^{(z,k)}\}$, which could eventually be used to build model assessment methods and/or to approach the conectivity problem between different brain regions. These issues are discussed in more detail in the Future Work section. The use of the \textbf{FEST} algorithm for the group case is straightforward and it just requiers the use of any of the distributions in (\ref{sec2:equ13}), (\ref{sec2:equ14}) or (\ref{sec2:equ15}) playing the role of $p(\theta^{(z)}_{vt,l}|D_{vt}^{(z)})$ and an appropriate estimation for observational noise variance $\mathbf{\Sigma}_{vt}^{(z)}$, which in this case can be computed as $\boldsymbol{\bar{S}}_{\scaleto{vt\mathstrut}{6pt}}^{(g)}=\frac{1}{N_{g}^2}\sum \limits_{z=1}^{n_g}\boldsymbol{S}_{\scaleto{vt\mathstrut}{6pt}}^{(z)}$. However, it requieres that the covariates related to $\mathbf{F}^{'(z)}_{vt}$ must be the same for all the subjects. This last is a common feature in many fMRI experiments as we will see in one of the examples presented below. However, some designs such as event-related designs occasionally requiere random sequences of stimuli, which lead to have different $\mathbf{F}^{'(z)}_{vt}$ matrices for each subject in the sample. To overcome this limitation, we propose a second and simpler algorithm, which does not depend on $\mathbf{F}^{'(z)}_{vt}$ and can also be used to perform inference about the state parameter.

\subsection{FSTS algorithm}

The evolution equation in (\ref{sec2:equ1}) relates to the state parameter at time $t$ with its own value at time $t-1$. We take advantage of that Markovian property, some prior-posterior estimations and the concept of discount factors to develop a forward sampler algorithm, which we called Forward State Trajectories Sampler (FSTS).

\begin{algorithm}
\caption{Forward State Trajectories Sampler}\label{euclid2}
\begin{algorithmic}[1]
\Procedure{$ \textsf{for}\ \ k=1\ldots N$}{}
\State Compute $\mathbf{W}_{vt}^{(z)} = \mathbf{B_t}\mathbf{C}_{v,t-1}^{(z)}\mathbf{B_t} - \mathbf{C}_{v,t-1}^{(z)}$ for $t=1,\ldots,T$
\State Draw $\mathbf{\Omega}_{vt}^{(z,k)}$ from $ N_{pq}[\mathbf{0}, \mathbf{W}_{vt}^{(z)}, \boldsymbol{S}_{\scaleto{vt\mathstrut}{6pt}}^{(z)}]$ for $t=1,\ldots,T$
\State Draw $\mathbf{\Theta}_{v,t-1}^{(z,k)}$ from $N_{pq}[\mathbf{m}_{v,t-1}^{(z)}, \mathbf{C}_{v,t-1}^{(z)}, \mathbf{S}_{v,t-1}^{(z)}]$ for $t=1,\ldots,T$
\State Compute $\mathbf{\Theta}_{vt}^{(z,k)}=\mathbf{\Theta}_{v,t-1}^{(z,k)}+\mathbf{\Omega}_{vt}^{(z,k)}$ for $t=1,\ldots,T$
\State Let $\hat{\mathbf{\Theta}}^{(z,k)}_{v} = \{ \mathbf{\Theta}_{v1}^{(z,k)}, \ldots, \mathbf{\Theta}_{vT}^{(z,k)}\}$

\EndProcedure
\end{algorithmic}
\end{algorithm}

From the vector $\hat{\mathbf{\Theta}}^{(z,k)}_{v} = \{ \mathbf{\Theta}_{v1}^{(z,k)}, \ldots, \mathbf{\Theta}_{vT}^{(z,k)}\}$, we can define three different types of sub-vectors related to the variables in (\ref{sec2:equ7}), (\ref{sec2:equ8}) and (\ref{sec2:equ9}). For instance, if one is interested in the joint effect (\ref{sec2:equ9}), then $\hat{\boldsymbol{\theta}}^{(z)}_{vl}=\{ \boldsymbol{\theta}_{v1,l}^{(z)}, \ldots, \boldsymbol{\theta}_{vT,l}^{(z)}\}$, for 
$l=1, \ldots, p$. Thus, for the task $l$ one can compute an activation probability for the voxel $v$ as

\begin{equation}\label{moteCarlo2}
p(\hat{\boldsymbol{\theta}}^{(z)}_{vl}>0) = E(\mathds{1}_{(\hat{\boldsymbol{\theta}}^{(z)}_{vl}>0)})\approx \frac{\sum\limits_{k=1}^{N} \mathds{1}_{(\hat{\boldsymbol{\theta}}^{(z,k)}_{vl}>0)}}{N}.
\end{equation}

The probability of activation in (\ref{moteCarlo2}) can be computed analogously for (\ref{sec2:equ7}) and (\ref{sec2:equ8}) just by taking the appropiate components from $\mathbf{\Theta}^{(z)}_{v}$.

\subsection{Forward-filtering-backward-sampling algoritm}

Following the same ideia as in \cite{fruhwirth1994data},  we now present a matrix-variate version of the forward-filtering-backward-sampling algorithm. It is worth mentioning that in \cite{fruhwirth1994data}, they use a data augmentation approach in order to tackle the problem related to the estimation of the covariance matrix $\mathbf{W}_{vt}^{(z)}$. Here, instead, we apply a discount factor approach as it is described above. Thus, using the Bayes' theorem and given some assumptions of conditional independence related to the MDLM, it can be shown that

\begin{equation*}
p(\mathbf{\Theta}_{v,t-j}^{(z)}|\mathbf{\Theta}_{v,t-j+1}^{(z)},\mathbf{\Sigma}^{(z)},D_{vt}^{(z)}) \propto p(\mathbf{\Theta}_{v,t-j}^{(z)}|\mathbf{\Sigma}^{(z)},D_{v,t-j}^{(z)})p(\mathbf{\Theta}_{v,t-j+1}^{(z)}|\mathbf{\Theta}_{v,t-j}^{(z)}, \mathbf{\Sigma}^{(z)},D_{v,t-j}^{(z)}),
\end{equation*}

which leads to

\begin{equation*}
(\mathbf{\Theta}_{v,t-j}^{(z)}|\mathbf{\Theta}_{v,t-j+1}^{(z)},\mathbf{\Sigma}^{(z)},D_{vt}^{(z)}) \sim N_{pq}(\mathbf{m}_{vj}^{*(z)}, \mathbf{C}_{vj}^{*(z)}, \mathbf{\Sigma}^{(z)}),
\end{equation*}

where $\mathbf{m}_{vj}^{*(z)} = \mathbf{m}_{v,t-j}^{(z)} + \mathbf{C}_{v,t-j}^{(z)}(\mathbf{B_t}\mathbf{C}_{v,t-j}^{(z)}\mathbf{B_t})^{-1}(\mathbf{\Theta}_{v,t-j+1}^{(z)} - \mathbf{m}_{v,t-j}^{(z)})$ and $\mathbf{C}_{vj}^{*(z)}=\mathbf{C}_{v,t-j} - \mathbf{C}_{v,t-j}(\mathbf{B_t}\mathbf{C}_{v,t-j}^{(z)}\mathbf{B_t})^{-1}\mathbf{C}_{v,t-j}$. Thus, the forward-filtering-backward-sampling algorithm for the matrix-variate case is given by:

\begin{algorithm}
\caption{Forward-filtering-backward-sampling}\label{euclid2}
\begin{algorithmic}[1]
\Procedure{$ \textsf{for}\ \ k=1\ldots N$}{}
\State Draw $\mathbf{\Sigma}^{(z,k)}$ from $W^{-1}_{n_t}(\mathbf{S}_{t})$ 
\State For $j=0$ draw $\mathbf{\Theta}_{v,t}^{(z,k)}$ from $N_{pq}(\mathbf{m}_{vj}^{(z)}, \mathbf{C}_{vj}^{(z)}, \mathbf{\Sigma}^{(z,k)})$
\State For $1\leq j<t$ draw $\mathbf{\Theta}_{v,t-j}^{(z,k)}$ from $(\mathbf{\Theta}_{v,t-j}^{(z)}|\mathbf{\Theta}_{v,t-j+1}^{(z,k)},\mathbf{\Sigma}^{(z,k)},D_{vt}^{(z)})$
\State Let $\hat{\mathbf{\Theta}}^{(z,k)}_{v} = \{ \mathbf{\Theta}_{v1}^{(z,k)}, \ldots, \mathbf{\Theta}_{vt}^{(z,k)}\}$
\EndProcedure
\end{algorithmic}
\end{algorithm}

Using Monte Carlo integration as in (\ref{moteCarlo2}), one can test brain activation just taking the appropiate components from $\mathbf{\Theta}^{(z)}_{v}$.

\section{Examples}
\label{sec:4}

In this section we present two applications modeling real fMRI data using model (\ref{sec2:equ1}) and applying the three algorithms presented above in order to detect brain reaction. In the first example, \cite{pernet2015human} perform an expriment where an auditoy stimulus, following a block-design, was presented in order to localize "temporal voice areas". In that work, \cite{pernet2015human} model the voice and non-voice sounds separate (separately), whereas here, for the sake of simplicity, we merge both vocal and non-vocal sounds in a single block design as it is shown in figure \ref{fig1}. In the second example, \cite{bazan2015motor} perform a finger-tapping task experiment, following an event-related design, which should lead to an activation of the motor cortex. With the aim of showing the performance of the three algorithms proposed in this work, we present the analysis for two single voxels (an active and non-active voxel) and activation maps for the entiere brain. In both examples, we run an analysis for one single subject and use the same fix value for the discount factor, i.e. $\beta_{it}=0.95$ and vague priors (zero mean and large variance) for the state parameter $\mathbf{\Theta}^{(z)}_{v0}$, as well as for $\mathbf{\Sigma}^{(z)}_{v}$ with $\boldsymbol{S}_{vo}^{(z)}=\boldsymbol{I}_q$ and $n_0=1$.

\subsection*{Auditory cortex activation}

In figure \ref{fig2}, we can see the results after applying the FAST, FSTS and FFBS algorithms to both active and non-active voxels, which are shown in figure \ref{fig1}. Considering that this is an experiment where only sound stimulation is applied, all of the three algorithms yield the results that are expected for voxels located at the temporal cortex and for voxels outside that region.

\begin{table}[H]
\begin{figure}[H]
  \centering
\begin{center}
\begin{tabular}{cc}
\multicolumn{2}{c}{\includegraphics[width=.50\textwidth]{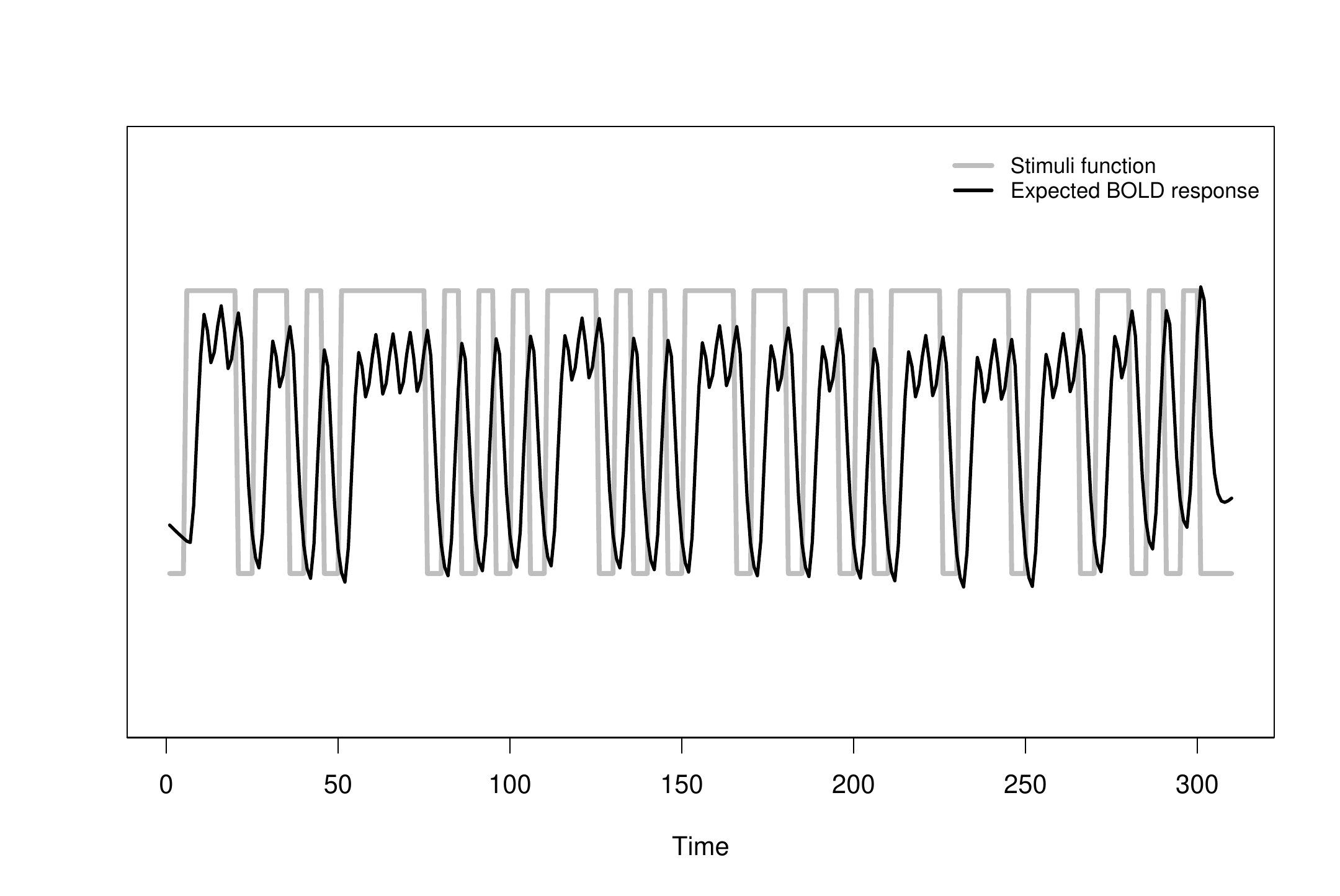}}\\
\includegraphics[width=.50\textwidth]{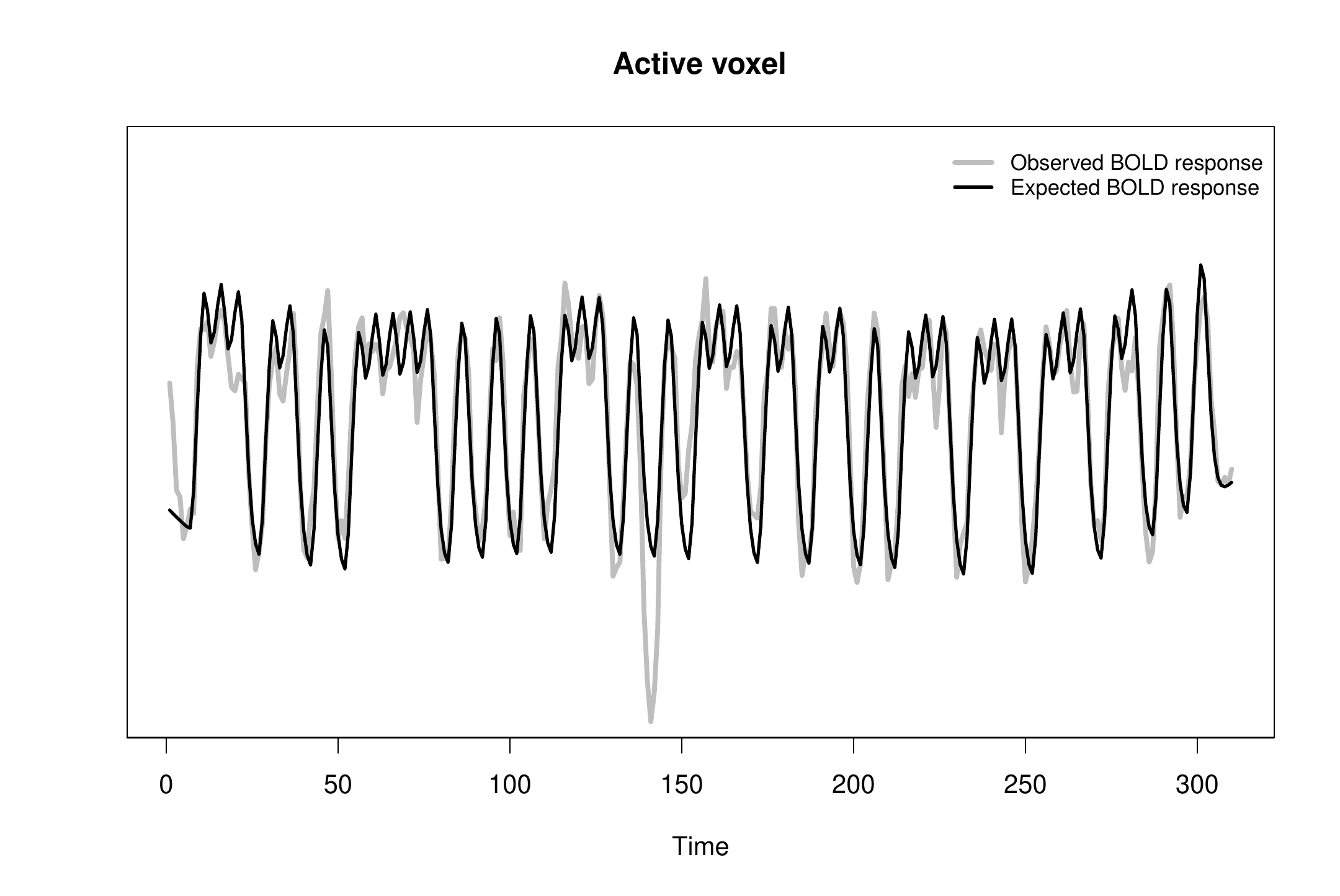}&\includegraphics[width=.50\textwidth]{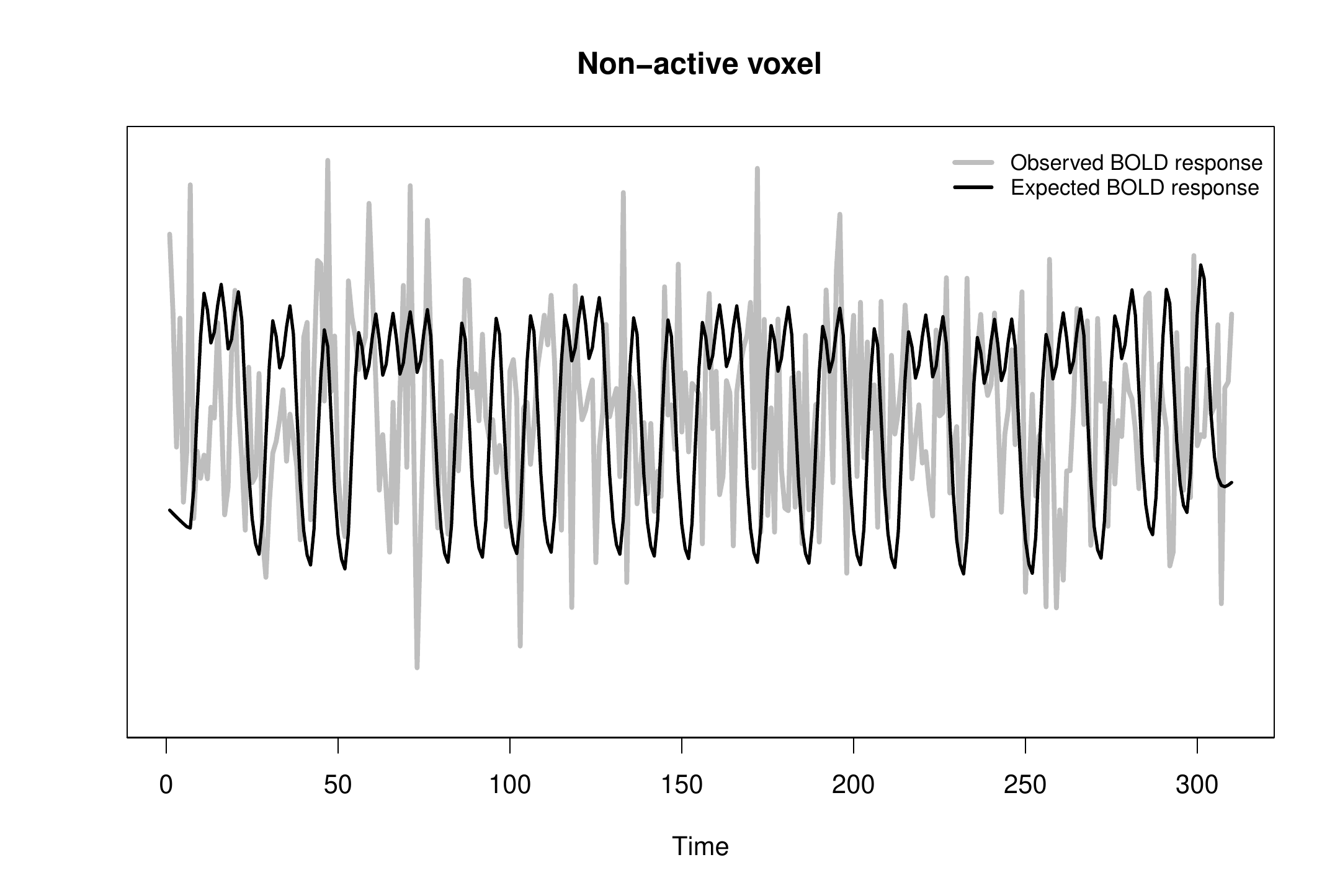}\\
\end{tabular}
\end{center}
  \caption{The top figure shows the block-design or stimuli function for voice and non-voice sounds and its expected Bold response. Lower left and right figures show the contrast (or possible match) between the expected and observed bold responses for both an active and non-active voxel respectively on a single subject.}
  \label{fig1} 
\end{figure}
\end{table}

\begin{table}[H]
\begin{figure}[H]
  \centering
\begin{center}
\begin{tabular}{cc}
Active Voxel & Non-active Voxel\\
\includegraphics[width=.50\textwidth]{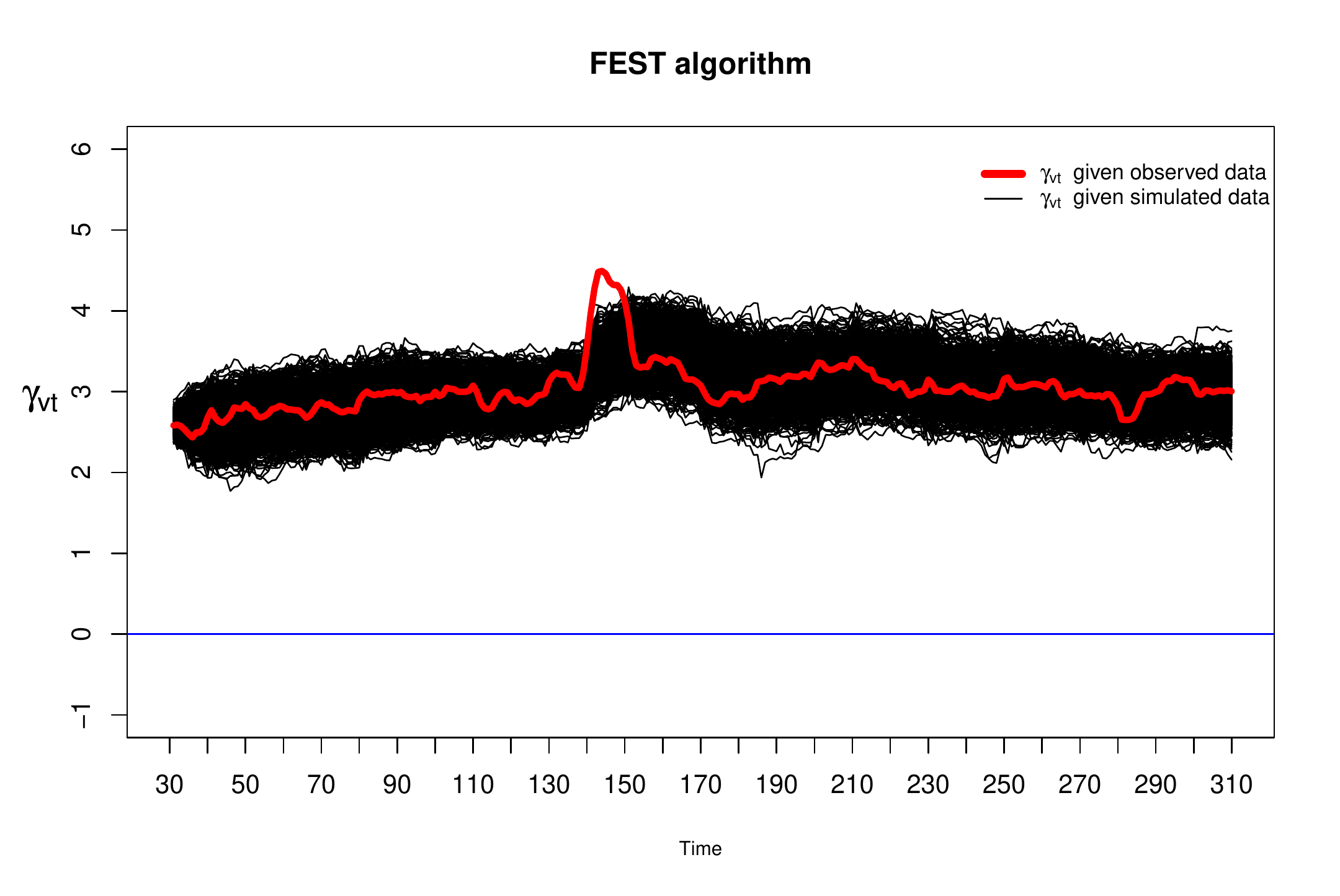}&\includegraphics[width=.50\textwidth]{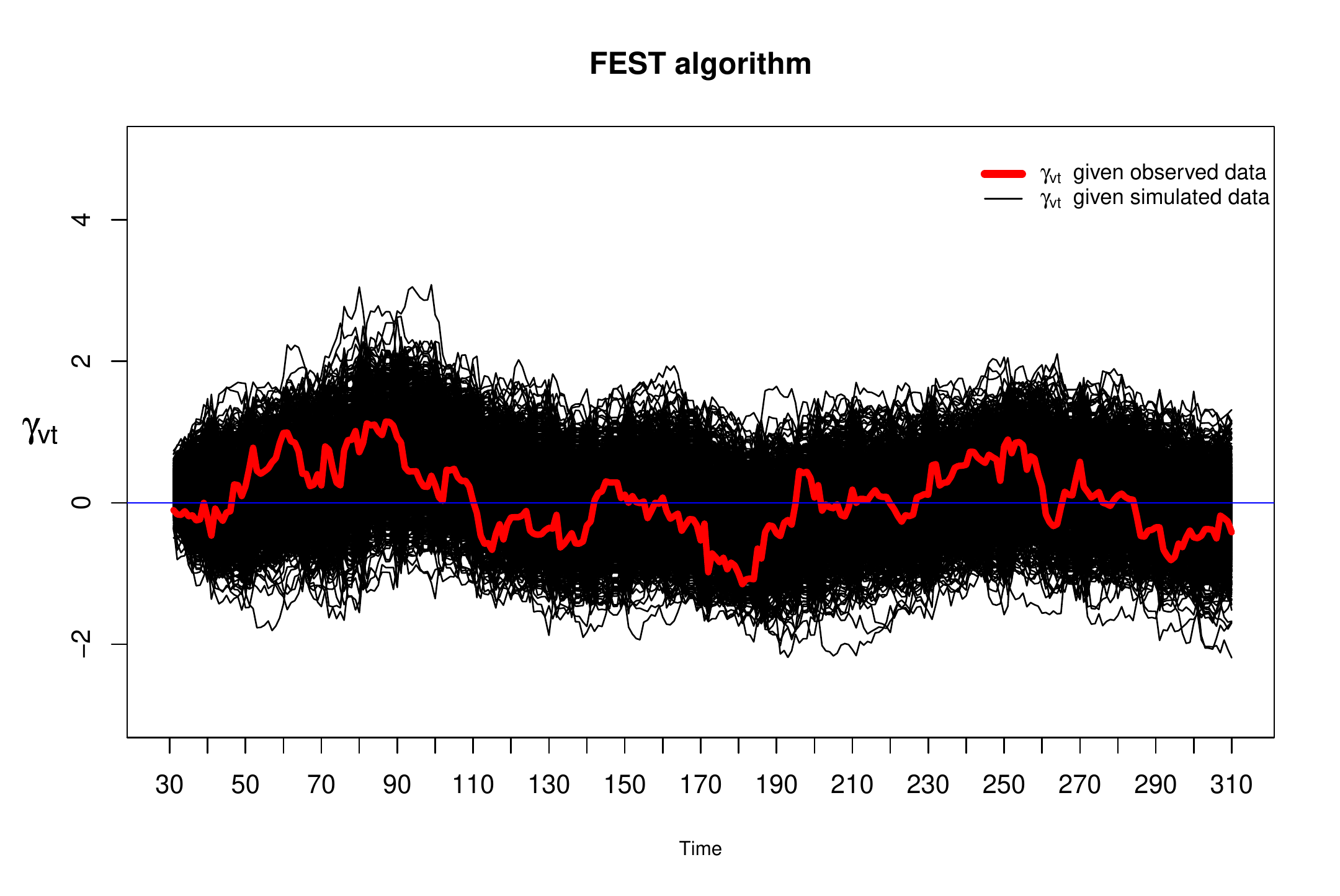}\\
\includegraphics[width=.50\textwidth]{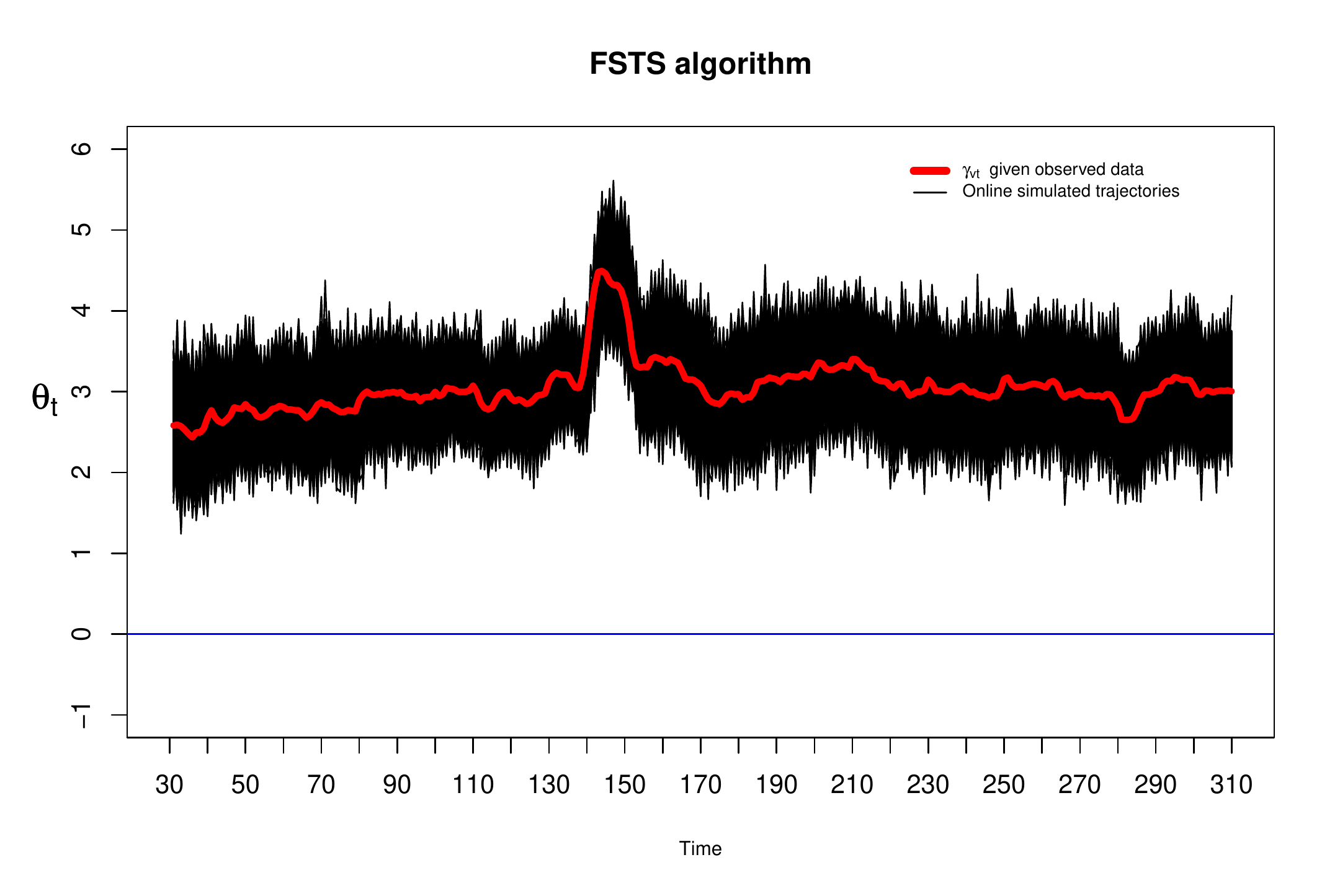}&\includegraphics[width=.50\textwidth]{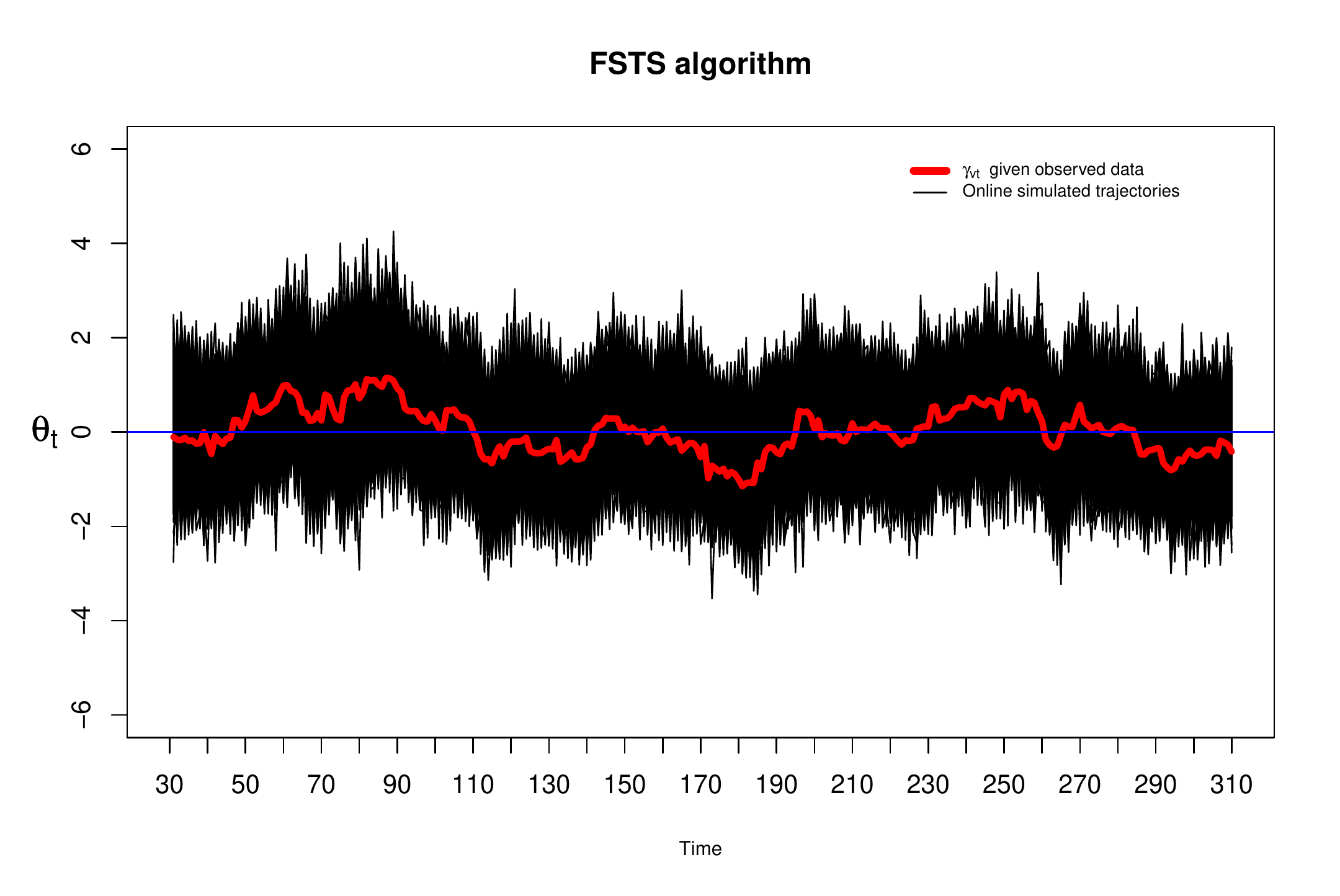}\\
\includegraphics[width=.50\textwidth]{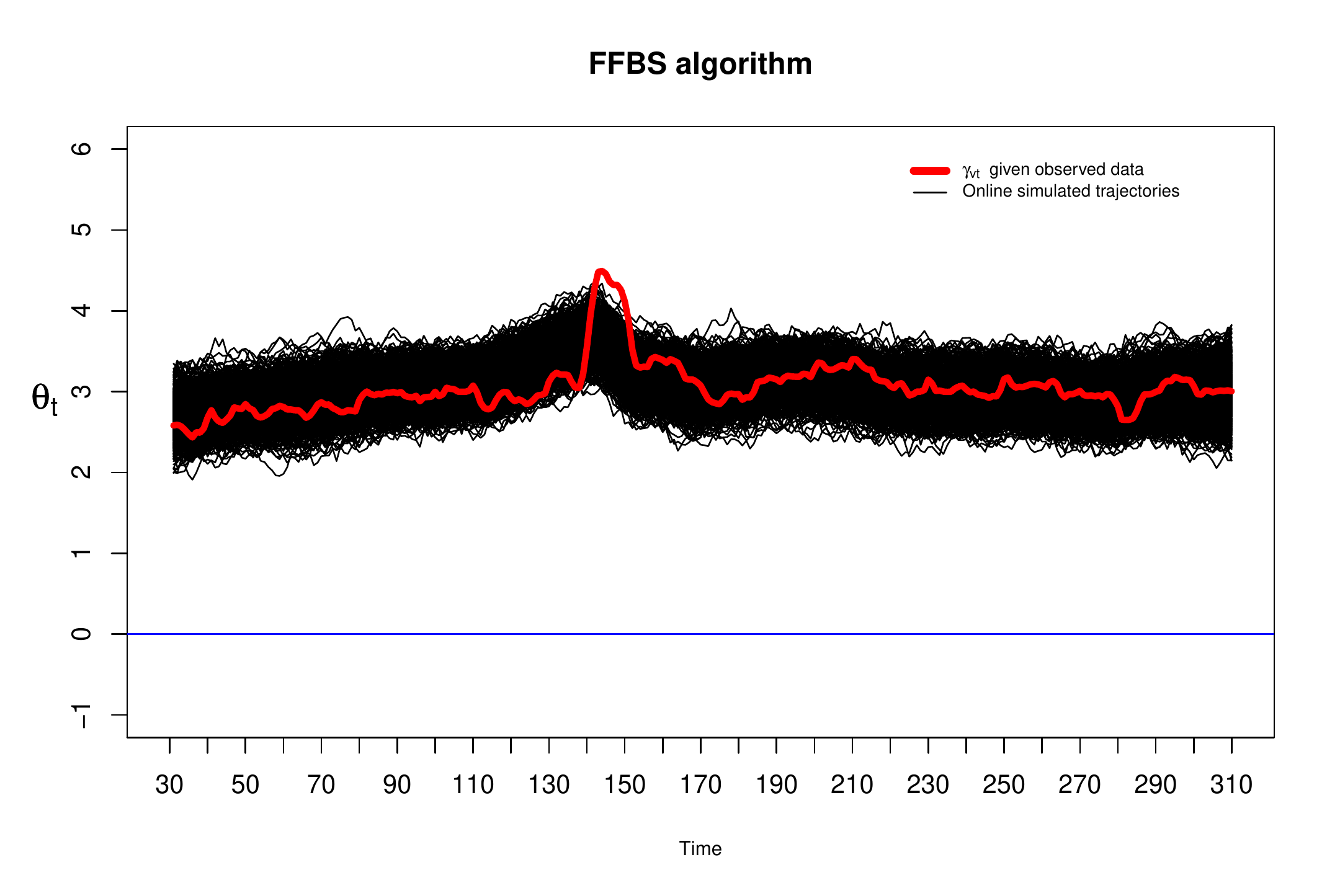}&\includegraphics[width=.50\textwidth]{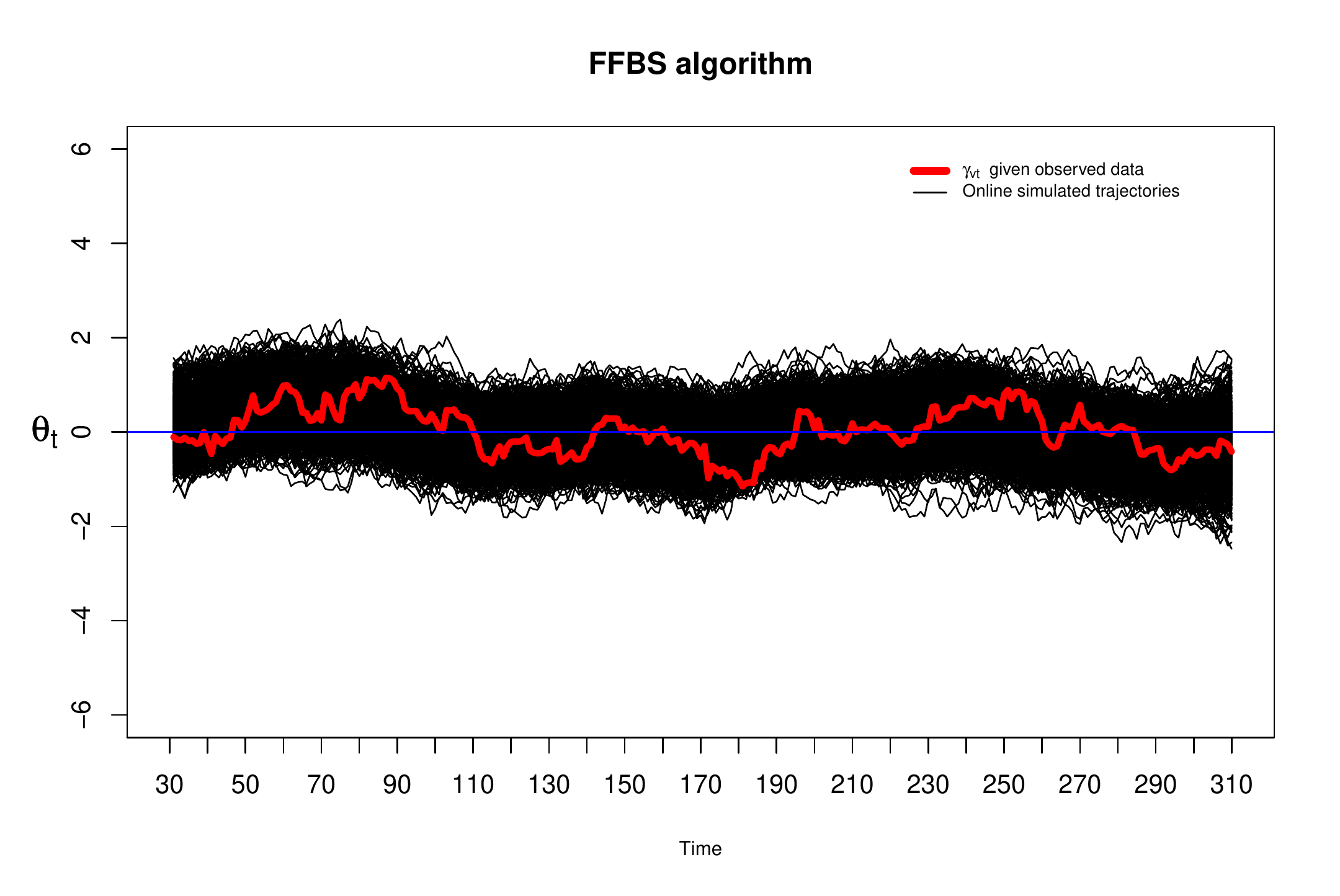}
\end{tabular}
\end{center}
  \caption{Simulated on-line trajectories obtained when  FEST, FSTS, and FFBS algorithms are applied to both active and non-active voxels presented on figure \ref{fig1}.}
  \label{fig2} 
\end{figure}
\end{table}

\vspace{-10mm}

For the active voxel, the on-line trajectories related to the state parameter lie above zero, which can be interpreted as a match between the expected and observed bold responses. For non-active voxels, the simulated on-line trajectories lie around zero, which yields lower evidence of activation (e.g. $<0.95$) according to the Monte Carlo evidence defined on (\ref{moteCarlo1}) and/or (\ref{moteCarlo2}). In figure \ref{fig3}, we present the activation maps for a single subject, obtained when applying the FEST, FSTS and FFBS algorithms respectively. For this example, we consider a voxel as active when its computed evidence obtained under any of the three algorithms, is greater than $0.95$. We also present a parametric map obtained when using the General Linear Model (with correction for multiple comparisons using Gaussian random-field theory and a significant level of 0.01\%), which can be considered the most common approach employed by users of statistical techniques for fMRI data analysis.

\begin{table}[H]
\begin{figure}[H]
  \centering
\begin{center}
\begin{tabular}{cc}
GLM & FEST-ACE\\
\includegraphics[width=.25\textwidth, angle =-90]{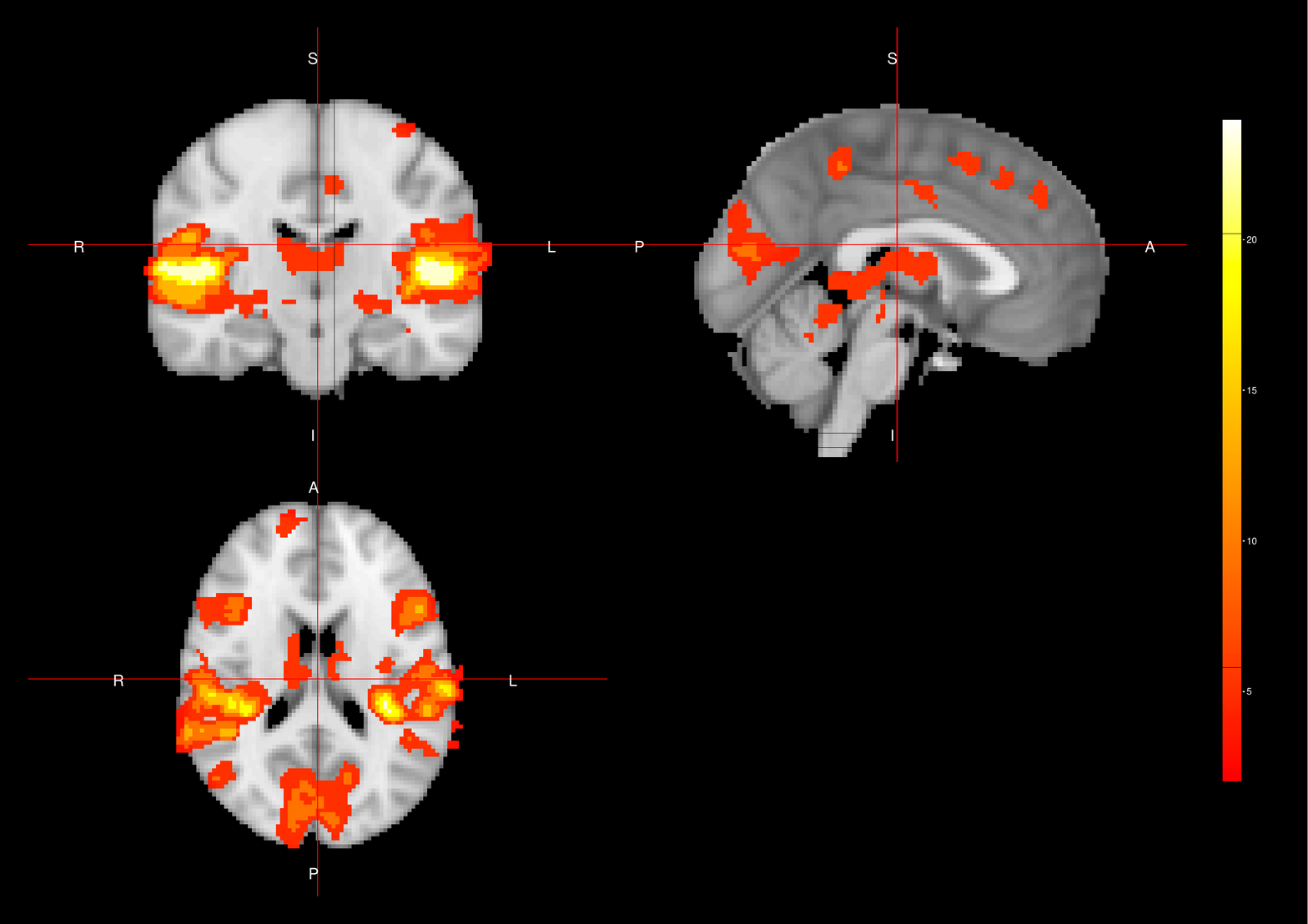}&\includegraphics[width=.25\textwidth, angle =-90]{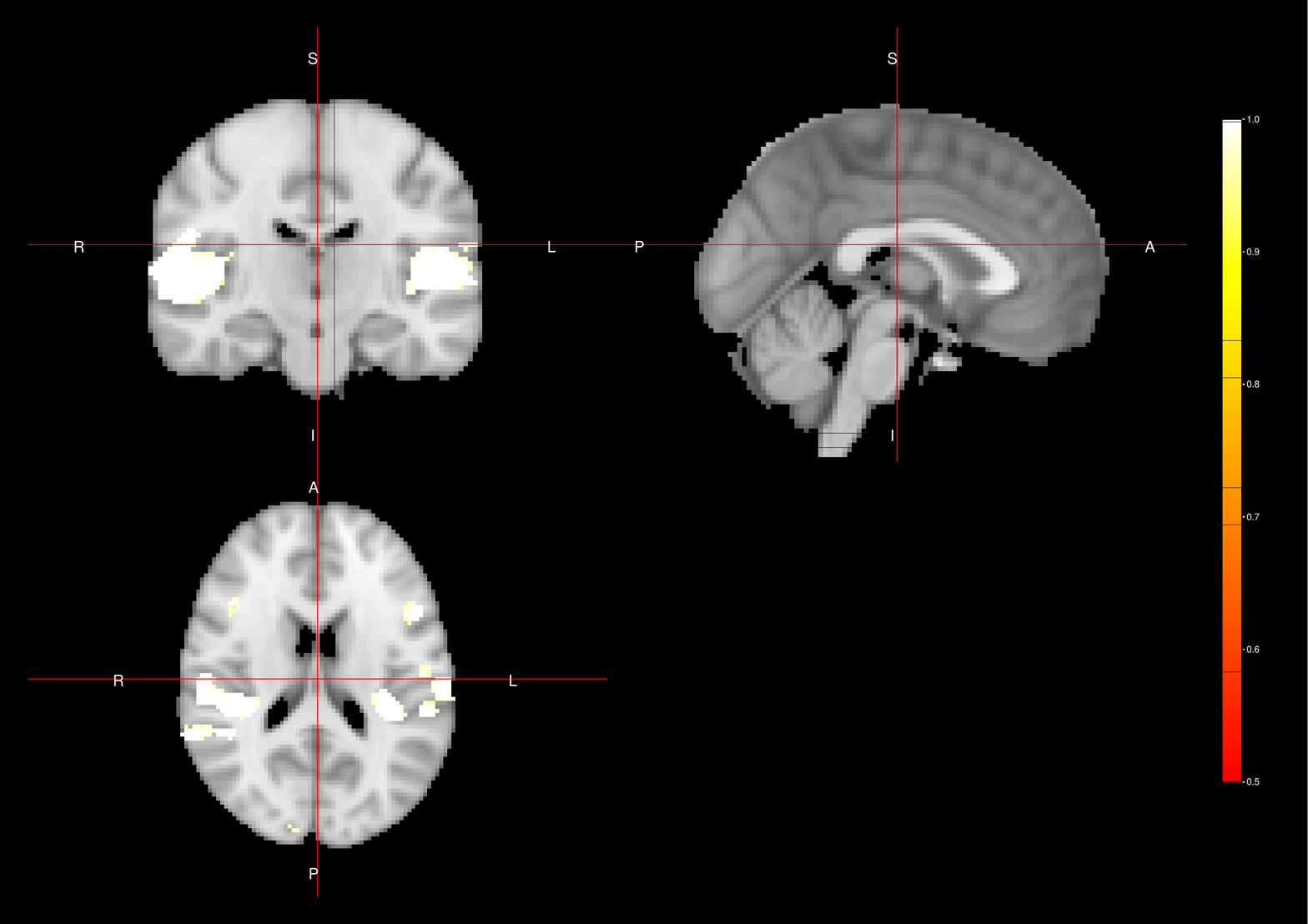}\\
&\\
FSTS-ACE & FFBS-ACE\\
\includegraphics[width=.25\textwidth, angle =-90]{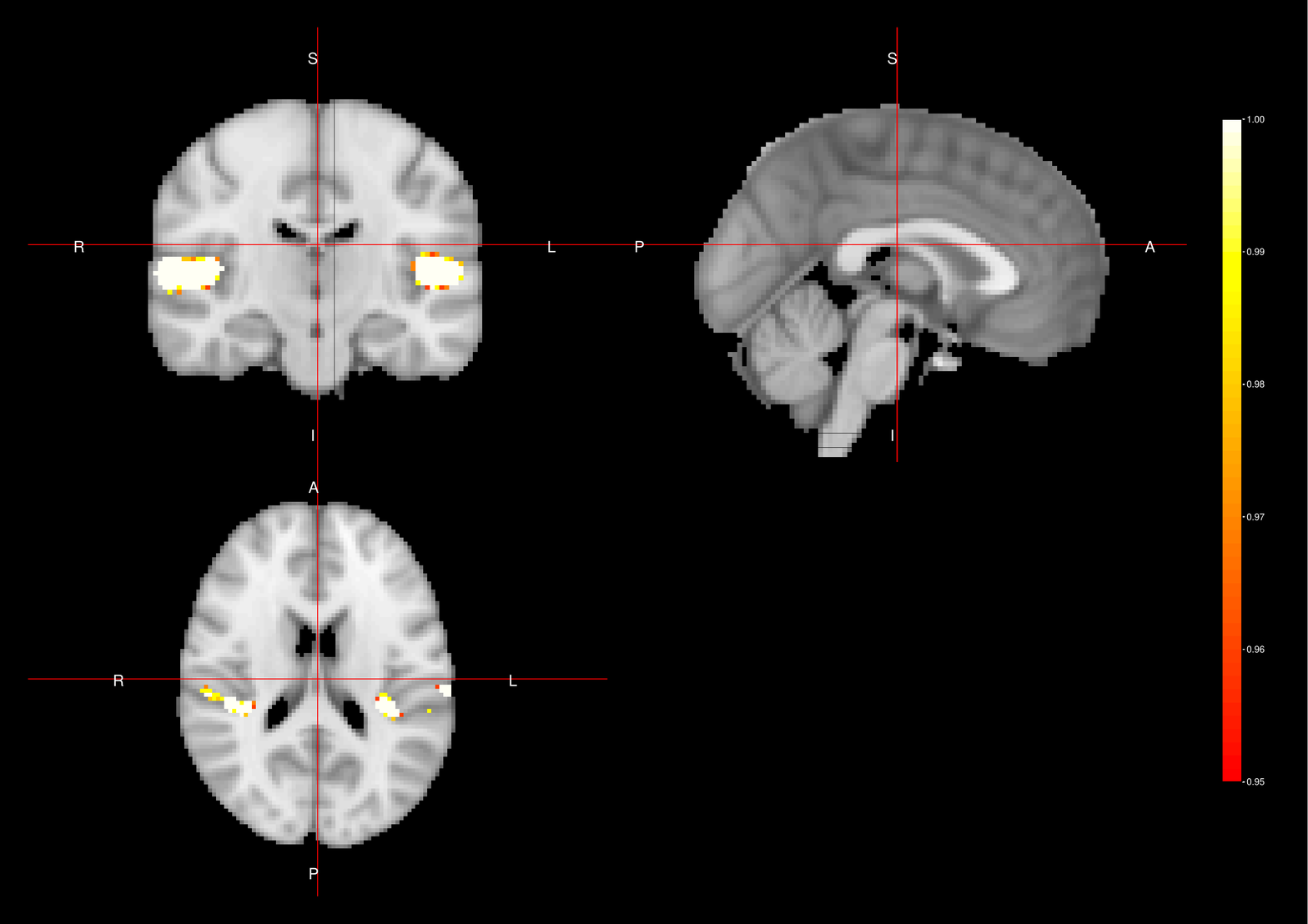}&
\includegraphics[width=.25\textwidth, angle =-90]{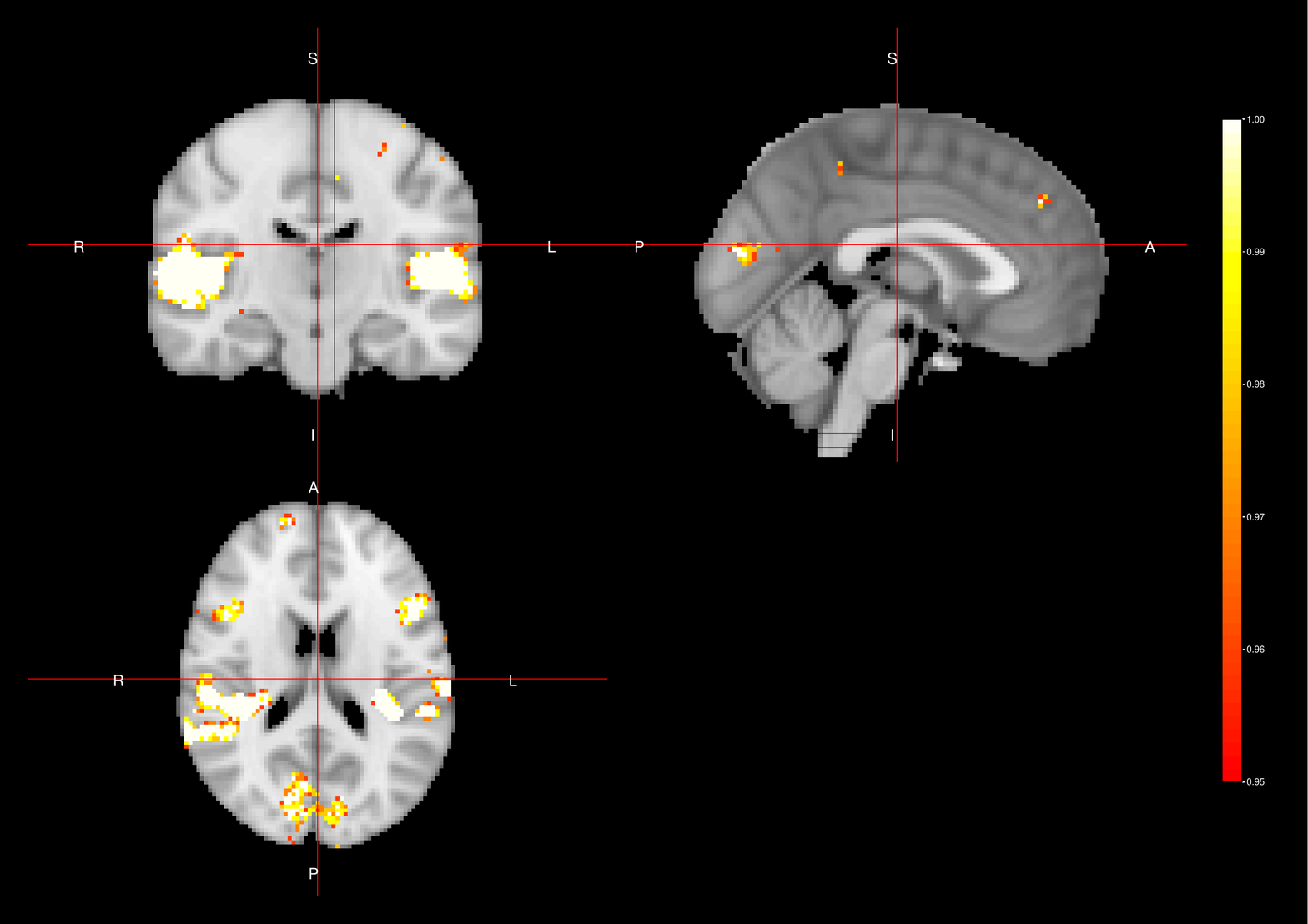}
\end{tabular}
\end{center}
  \caption{Voxel-wise analysis for one single subject. The top left figure shows a parametric map obtained via General Linear Model (GLM). The top right and lower left and right figures show activation maps related to each of the FEST, FSTS and FFBS algorithms respectively.}
  \label{fig3} 
\end{figure}
\end{table}

From figures \ref{fig2} and \ref{fig3}, we can see the performance of our method in this example, with each of the three algorithms detecting the parietal activation due to sound stimulation. From our knowledge and after evaluating these images with some colleagues from the field of Radiology, we conclude that the FEST algorithm yields better results in terms of a wider parietal activation and less ("false") activations from outside the temporal cortex. FFBS also yields good activation maps, but with more activations from outside the temporal cortex. FSTS also performs well, but shown to be conservative with almost no possible false activations detected and with a smaller parietal activation. For this example, only the average cluster effect (ACE) distribution (\ref{sec2:equ8}) is employed as the sampler distribution for the state parameter. However, the results are pretty similar when using either the marginal (\ref{sec2:equ7}) or the joint (\ref{sec2:equ9}) effect distribution.

\subsection*{Motor cortex activation}  

\begin{table}[H]
\begin{figure}[H]
  \centering
\begin{center}
\begin{tabular}{cccc}
$\beta=0.95$ & $\beta=0.90$&$\beta=0.87$ & $\beta=0.85$\\
\includegraphics[width=.2\textwidth]{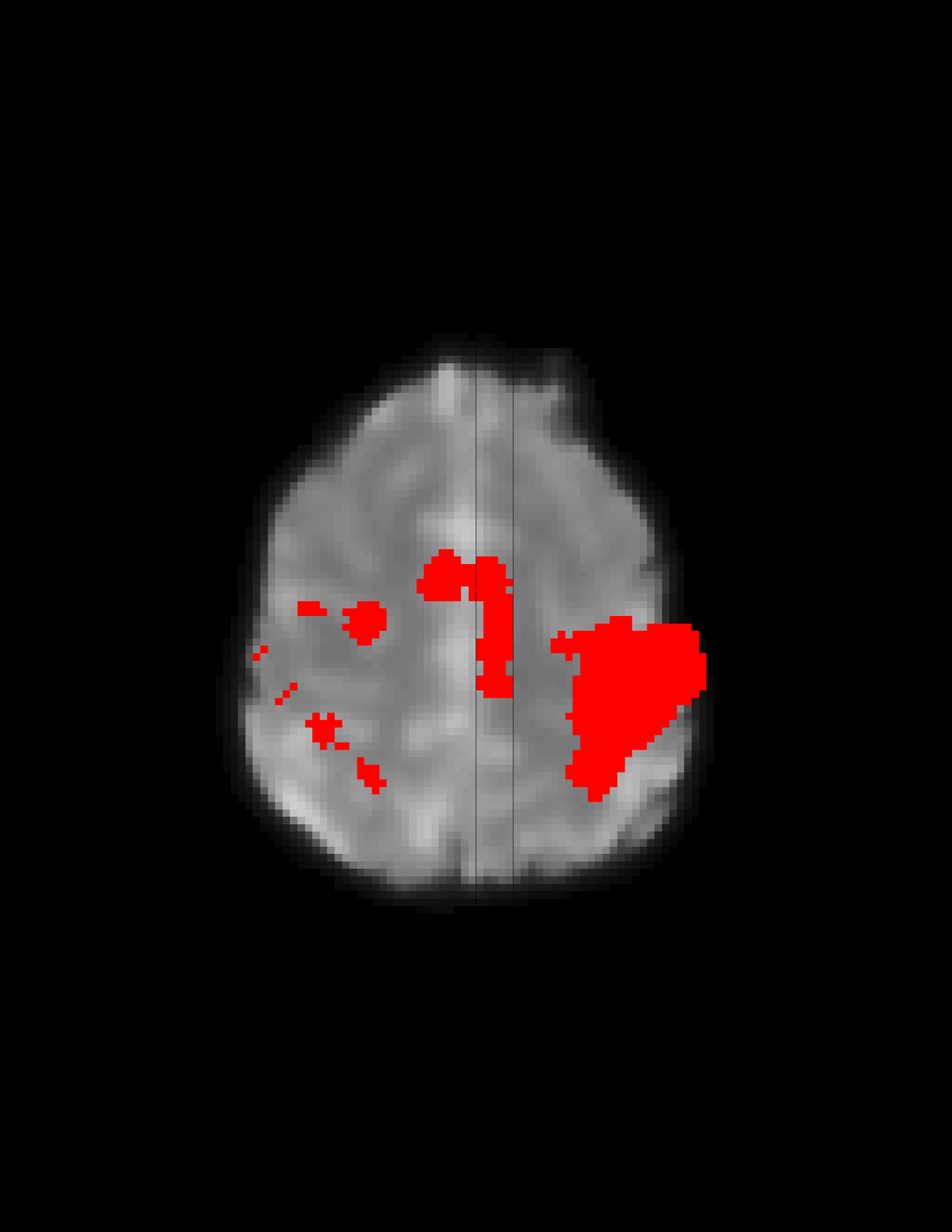}&\includegraphics[width=.2\textwidth]{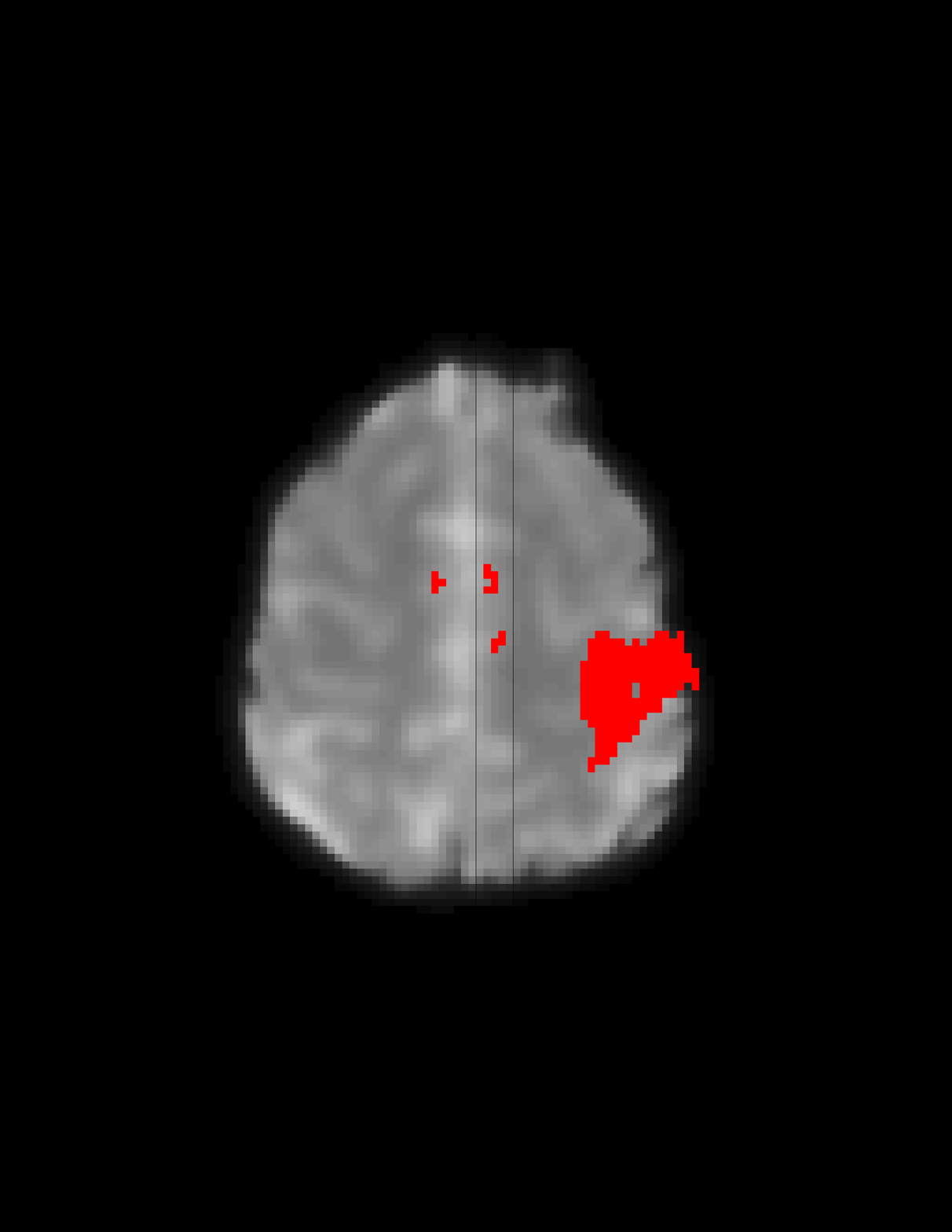}&
\includegraphics[width=.2\textwidth]{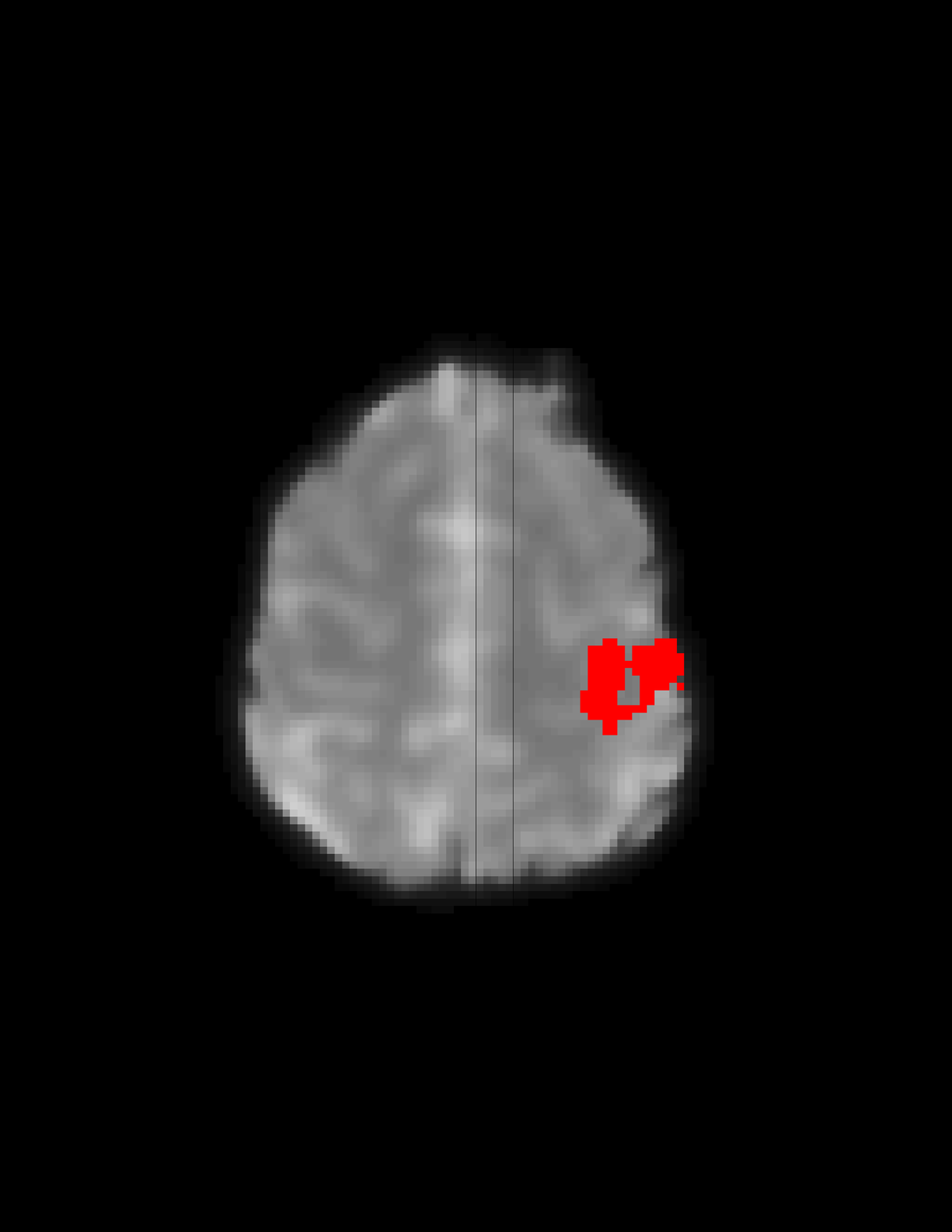}&
\includegraphics[width=.2\textwidth]{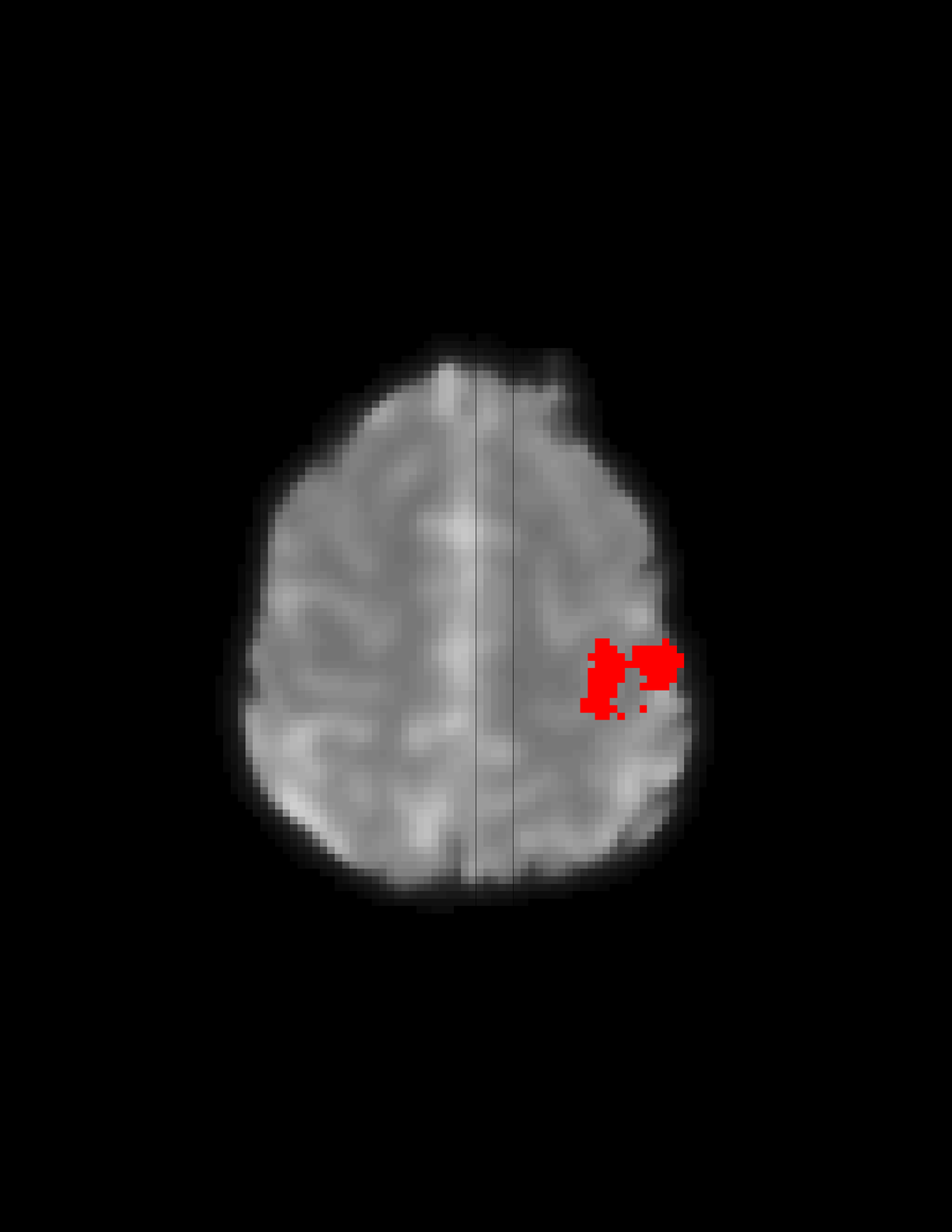}
\end{tabular}
\end{center}
  \caption{Activation maps of 95\% obtained using the FEST-ACE algorithm for different values of the discount factor $\beta$.}
  \label{fig4} 
\end{figure}
\end{table}

In figure \ref{fig4}, we can see an activation map corresponding to the event-related finger-tapping experiment from \cite{bazan2015motor} obtained when using the FEST algorithm. When the FSTS and FFBS algorithms are applied, they yield pretty similar results successfully identifying the motor cortex activation expected from this experiment. What we aiming to highlight with this example is the key role played by the discount parameter $\beta$. In the modeling we are proposing here, we let the value of $\beta$ as a free input value defined by the user. However, we can see that the activation map is sensitive to the value set for $\beta$. From our empirical experience, we recommend values for $0.8<\beta<1$.
One possible way to overcome this limitation is to run several maps for different $\beta$ values and use a statistical criterion (e.g. Bayes factor) to select the "appropriate" map. In the final section, we discuss this issue in more detail.

\section{Assessment of the Method}
\label{sec:5}

In order to assess the activation maps obtained when the model (\ref{sec2:equ1}) is fitting and any of the algorithms proposed in this work is used to detect voxel activation, we employ two different approaches. For the first type of assessment, we resort to simulated data using the R package \textbf{neuRosim} (\cite{welvaert2011neurosim}). The second assessment approach is inspired by the work of \cite{eklund2016cluster}, where real data from resting state experiments is used in order to evaluate the rate of false positive activations in the entire brain.

\subsection{Using simutaled data}

The aim of using simulated data is to evaluate the capacity of our method to detect true voxel activations, which in this case are artificially created using the R package \textbf{neuRosim}. Specifically, we create two spherical shape activations in the visual cortex related to a block design experiment.
In \cite{welvaert2011neurosim} the interested reader can find a detailed explanation on how to simulate fMRI data using the \textbf{neuRosim} package. In this aeessment, we set two different radius sizes (4 and 7 respectively) and the same effect size (250) for the two activated regions. In the figure \ref{fig5}, we can see the activation maps obtained when using each one from the three algorithms FEST, FSTS, and FFBS, for two different levels of signal-to-noise ratio (\textbf{SNR}).
From figure \ref{fig5}, we can see that when a high SNR (30) is fixed, all the three algorithms successfully identify the activated regions. However, when a low SNR (3.2) is set the FSTS algorithm fails to identify the activated regions, even when a less conservative probability threshold is applied (90\%).
The activation maps, when using FEST and FFBS algorithms under both scenarios yield quite good results, with a notorious performance of the FEST algorithm in terms of none false-positive activations.

\begin{table}[H]
\begin{figure}[H]
  \centering
\begin{center}
\begin{tabular}{ccc}
\multicolumn{3}{c}{\textbf{SNR}=30}\\
FEST (95\%) & FSTS (95\%) & FFBS (95\%)\\
\includegraphics[width=.1\textwidth]{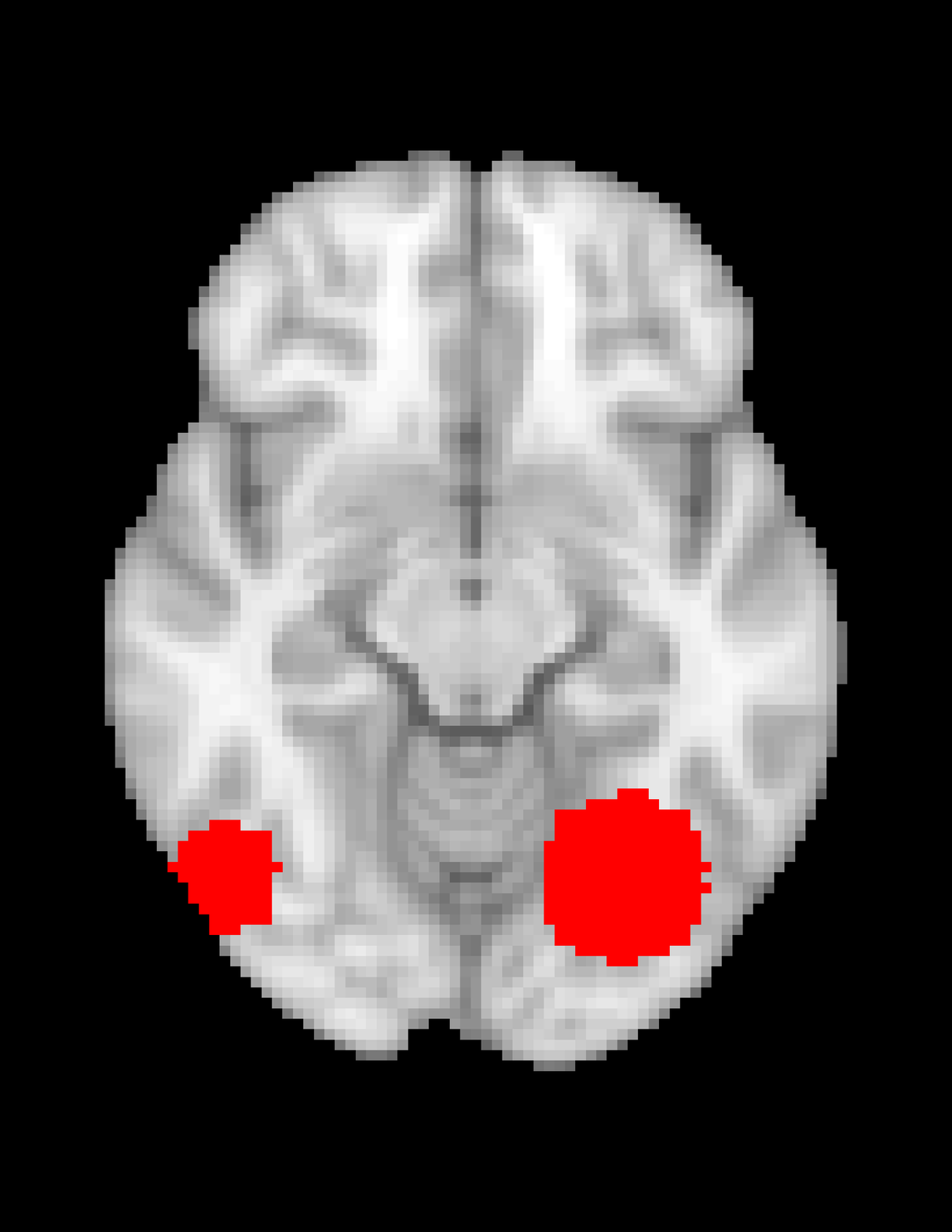}&\includegraphics[width=.1\textwidth]{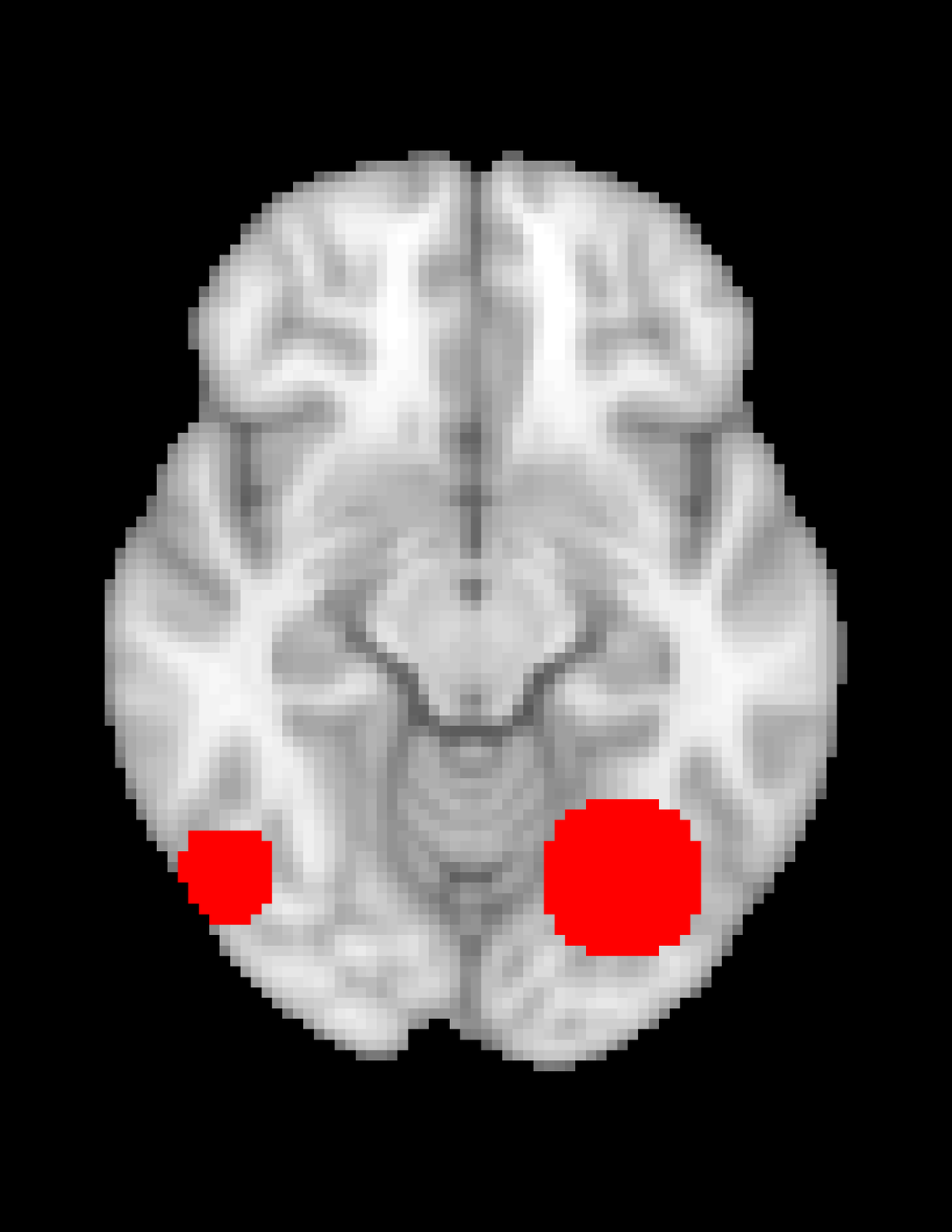}&\includegraphics[width=.1\textwidth]{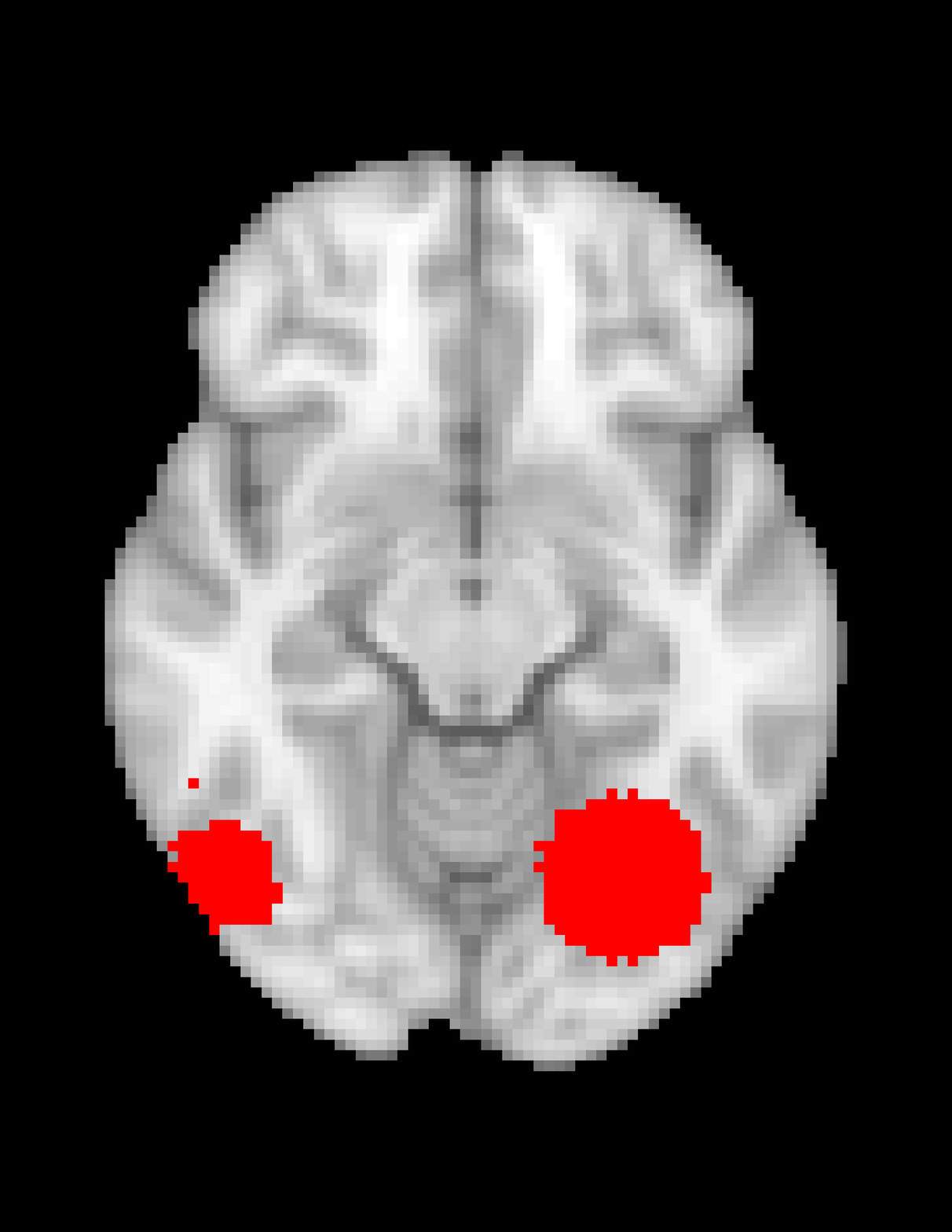}\\

\multicolumn{3}{c}{\textbf{SNR}=3.2}\\
FEST (95\%) & FSTS (95\%) & FFBS (95\%)\\
\includegraphics[width=.1\textwidth]{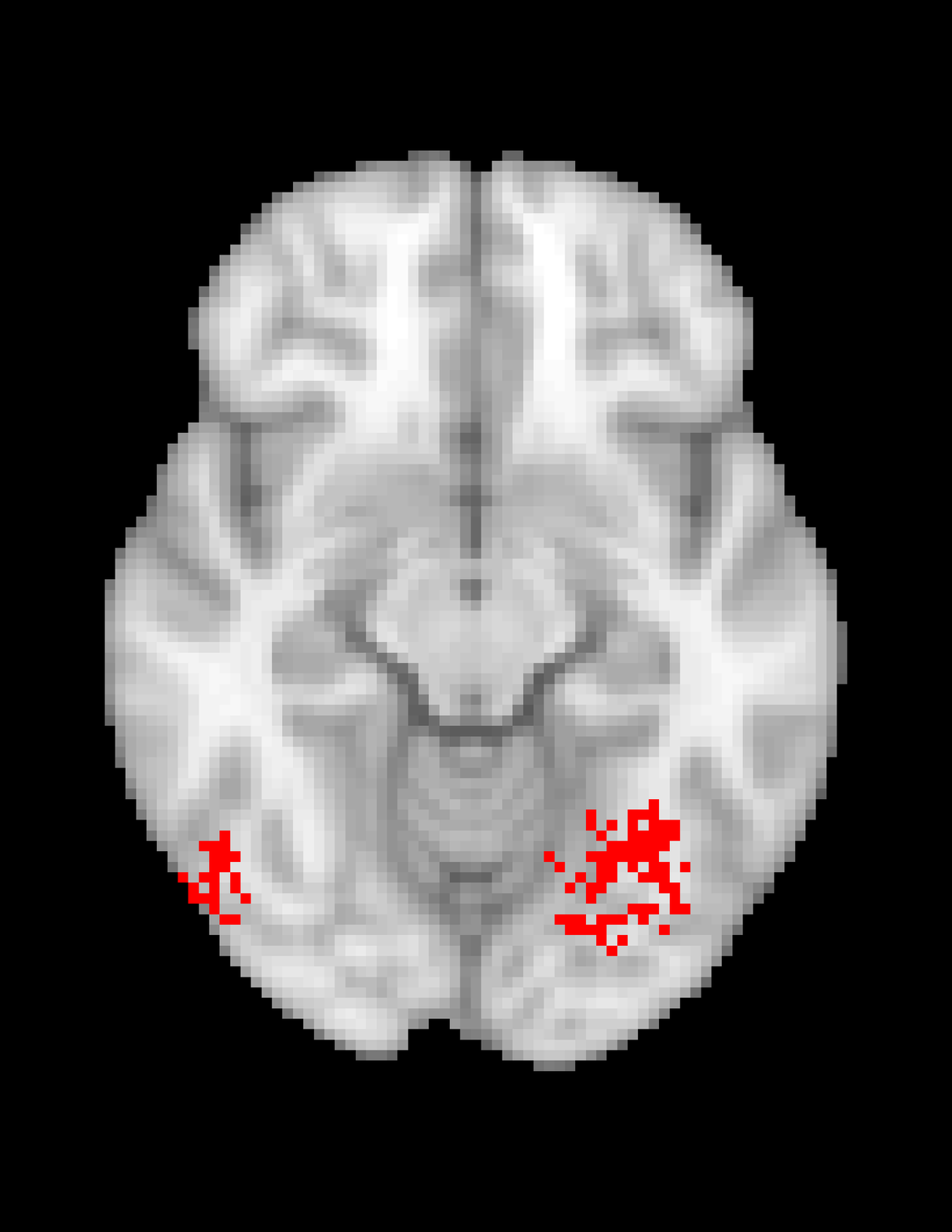}&\includegraphics[width=.1\textwidth]{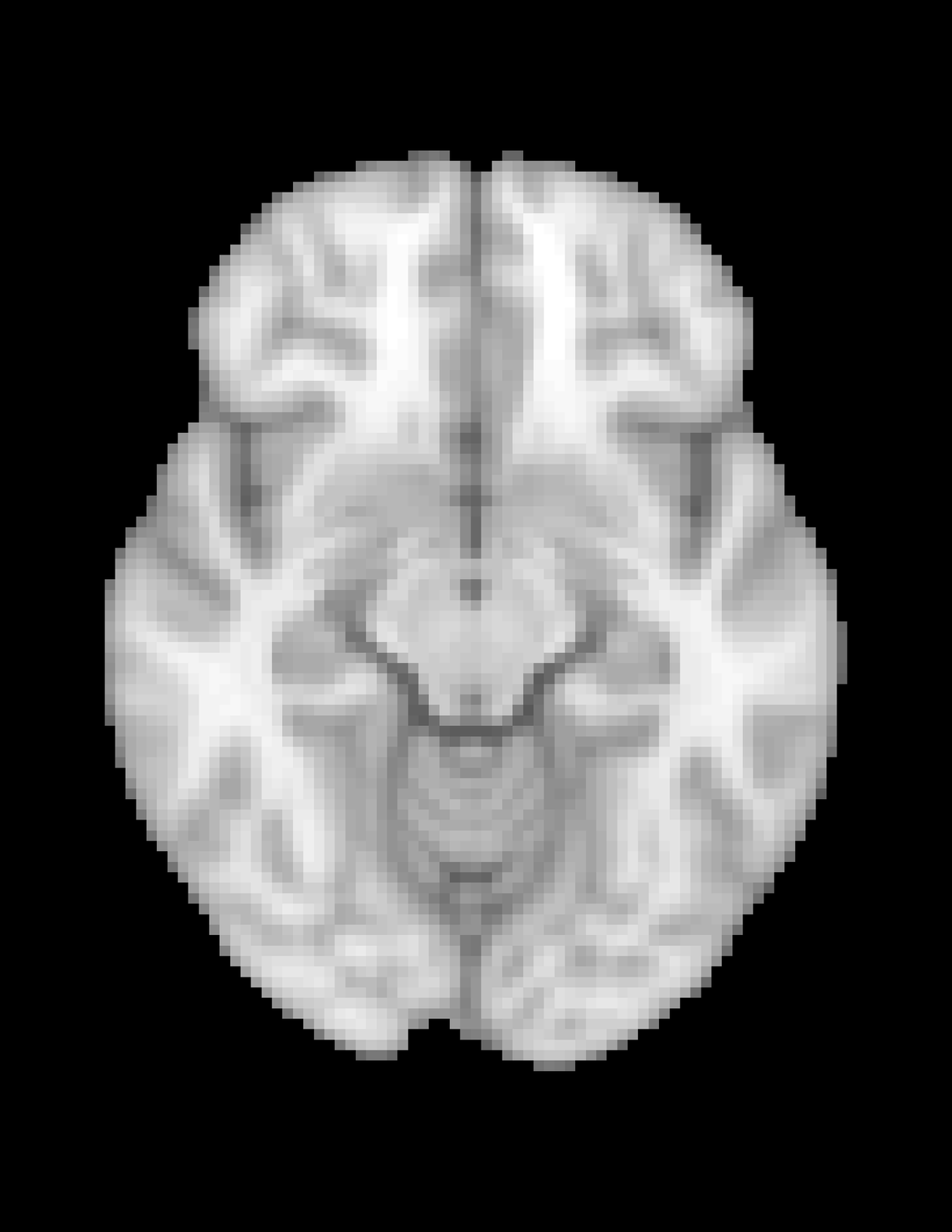}&\includegraphics[width=.1\textwidth]{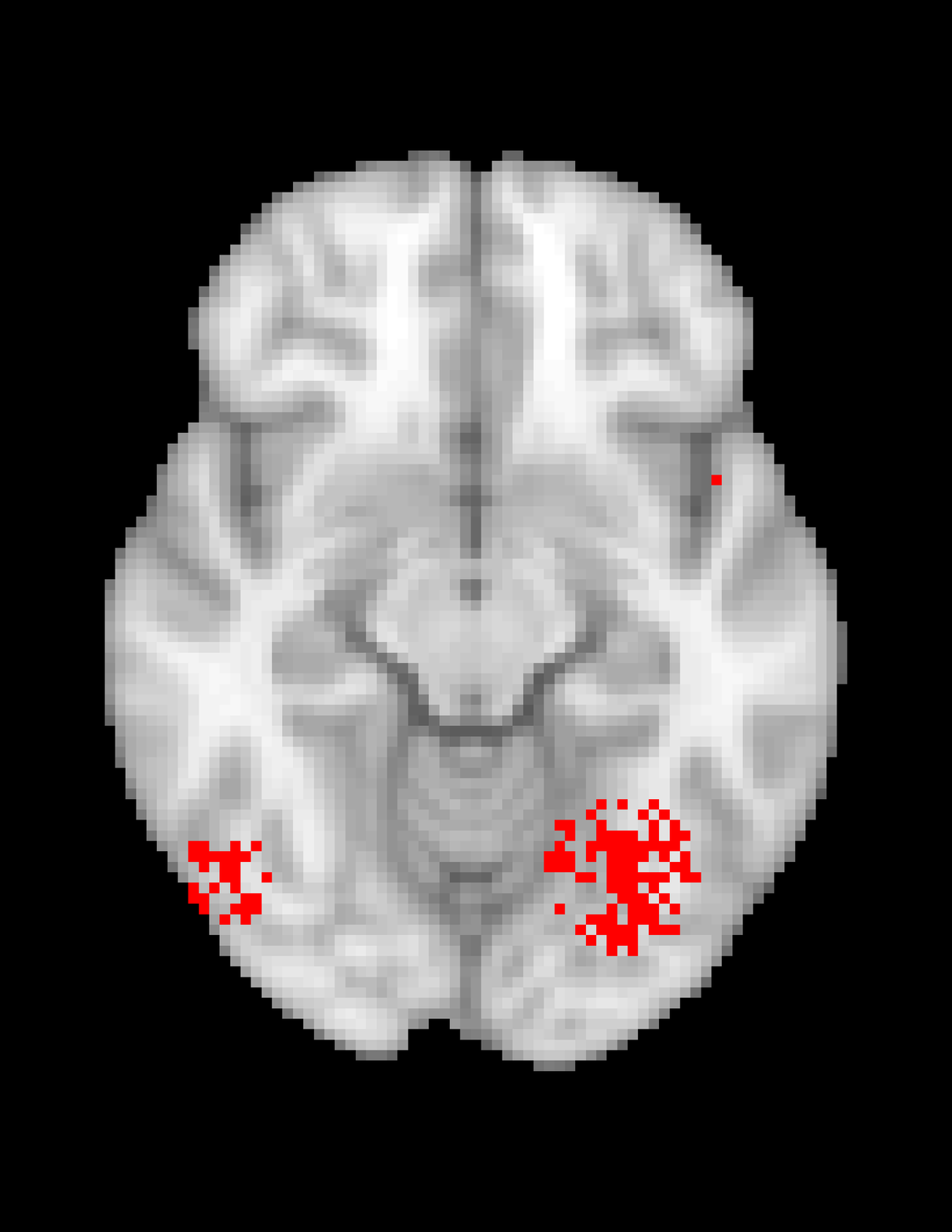}\\
FEST (90\%) & FSTS (90\%) & FFBS (90\%)\\
\includegraphics[width=.1\textwidth]{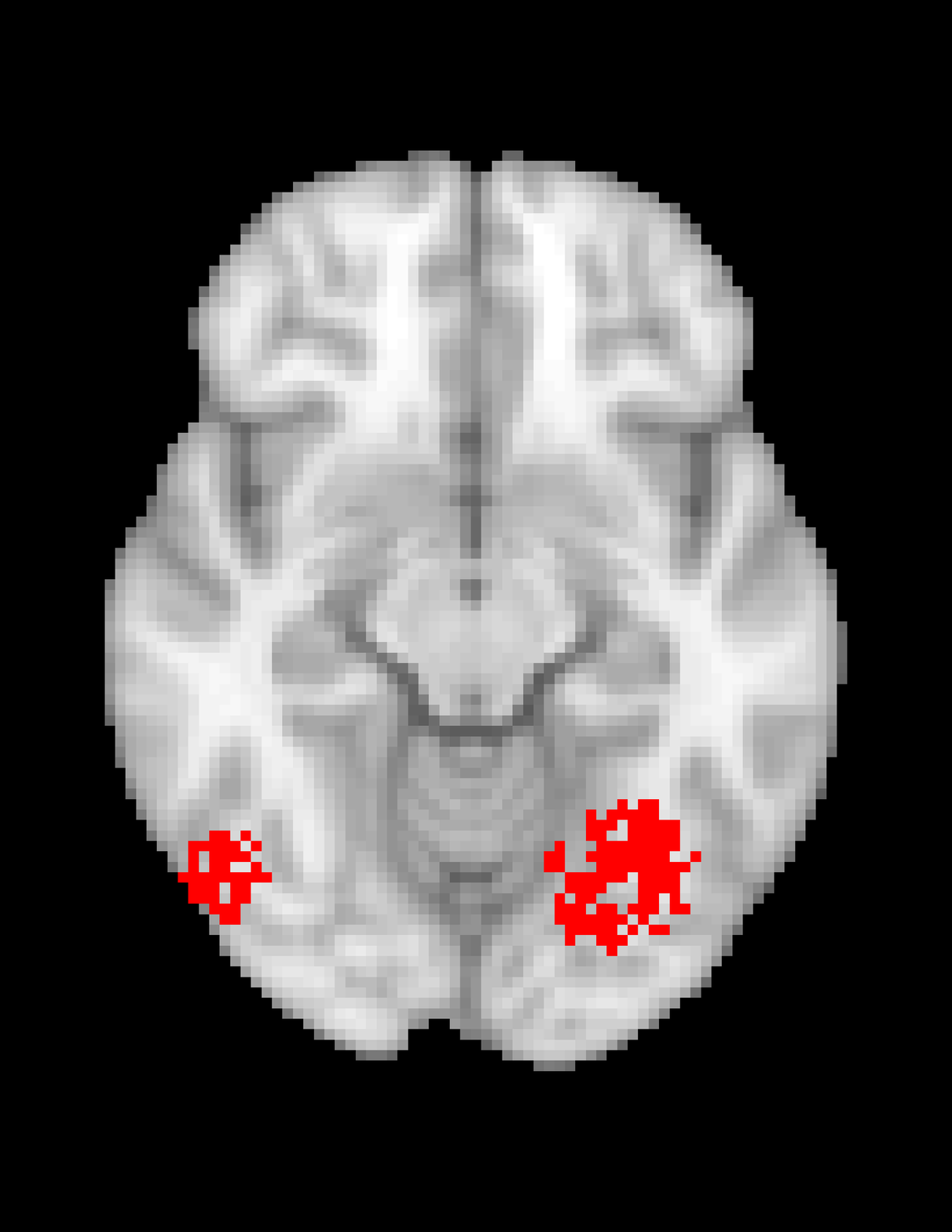}&\includegraphics[width=.1\textwidth]{simulatedDataFSTS_32_95.pdf}&\includegraphics[width=.1\textwidth]{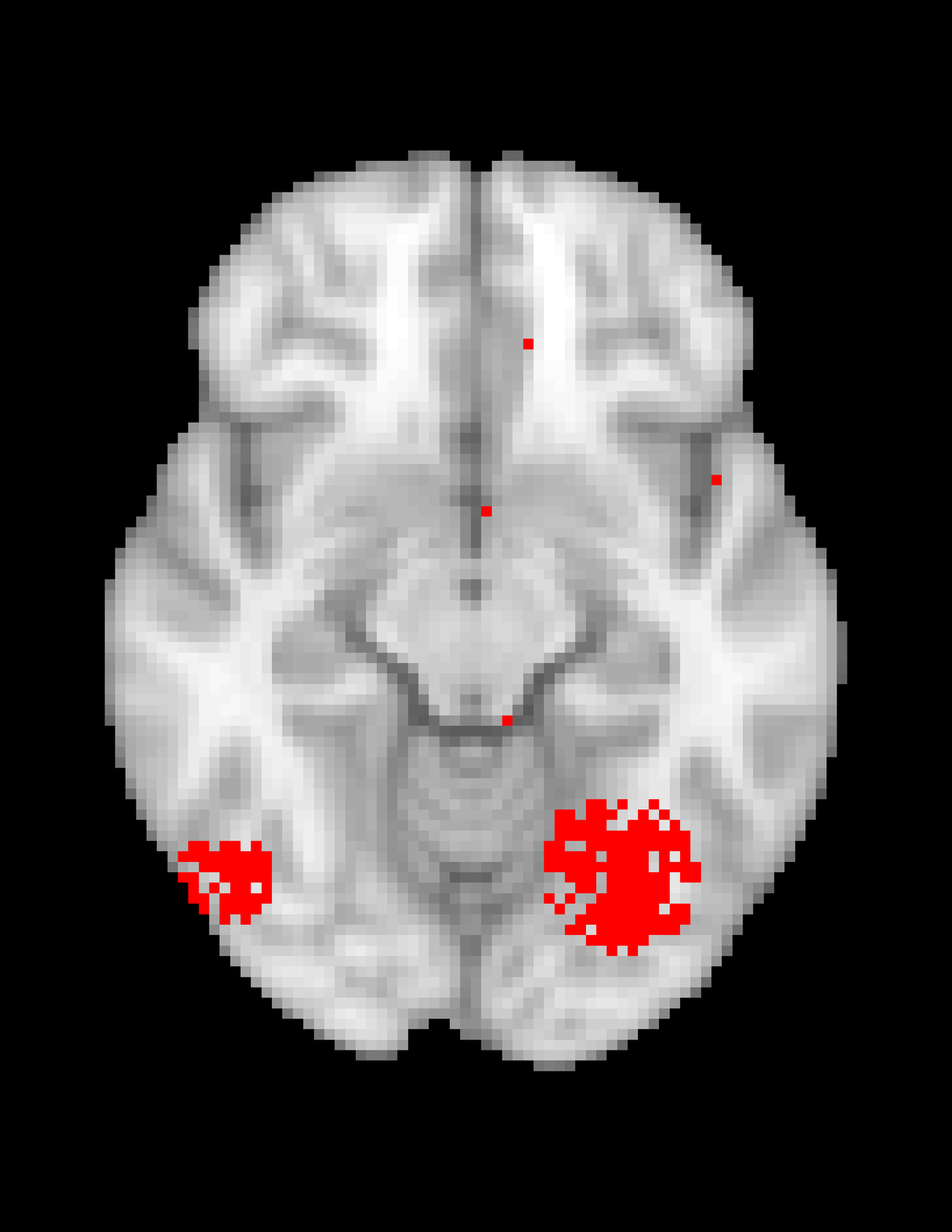}

\end{tabular}
\end{center}
  \caption{activation maps for fMRI simulated data under different \textbf{SNR} levels, algorithms (FEST, FFBS and FFBS), and probability activation thresholds.}
  \label{fig5} 
\end{figure}
\end{table}

\subsection{Using real data}

One of the big challenges to the statistical techniques for fMRI data analysis is controling the rate of false-positve activations, which usually appear on activation maps. For instance, in the case of the GLM, even using some sort of corrections for multiple comparisons, the rate of false-positive can be as high as 70\% (\cite{eklund2012does} and \cite{eklund2016cluster}). Thus, in order to assess the capacity of the method proposed in this work related to the control of false activations, we follow the same approach as in \cite{eklund2016cluster}. That approach consists on the use of real fMRI data, related to resting-state experiments obtained from the 1000 Functional Connectomes Project \cite{biswal2010toward}, which are correlated with covariates from fictitious stimuli. Hence, all the significant activations detected are interpreted as false-positive activations. This is a resaonable assumption in the sense that if one models the observed BOLD signal ($Y^{(z)}_{vt}$) from a resting-state experiment as $E(Y^{(z)}_{vt}) = \mathbf{F}^{'}_{vt}\mathbf{\Theta}_{vt}$, where $\mathbf{F}^{'}_{vt}$ is an (fictitious) expected BOLD signal not related with $Y^{(z)}_{vt}$, what one would expect to find is $E(\mathbf{\Theta}_{vt}|D_{vt}))\leq 0$, which can be interpreted as no existing match between the expected and observed BOLD signals. 
Therefore, we create four fictitious variables related to two block paradigms (B1 (10-s on off), B2 (30s on off)) and to two event-related paradigms (E1 (2-s activation, 6-rest), E2(1- to 4-s activation, 3- to 6s rest, randomized)). We correlate those variables with the fMRI resting data from three different labs (Beijing(198 subjects), Cambridge (198 subjects), Oulu (103 subjects)), and count the number of (false) significant activations. This analysis is only performed at the individual level.

\begin{table}[H]
  \begin{center}
    \caption{Rate of false-positive activations for the fictitious block paradigms B1 and B2.}
    \label{tab:table1}
\begin{tiny}
\begin{tabular}{|l|c|c|c|c|c|c|c|c|c|}
\hline
\multicolumn{10}{|c|}{B1}\\
\hline
 &\multicolumn{3}{c|}{FEST}&\multicolumn{3}{c|}{FSTS}& \multicolumn{3}{c|}{FFBS}\\
\cline{2-10}
Source fMRI data     &   Marginal  & LTT & Joint&  Marginal & LTT & Joint&  Marginal  & LTT & Joint\\
\hline
Beijing (198 subjects)    &0.0 & 0.0& 0.0&0.0 & 0.0&0.0 &2.5$\times 10^{-3}$ &4.1$\times 10^{-3}$&6.7$\times 10^{-6}$ \\
Cambridge (198 subjects)  &7.9$\times 10^{-5}$ &1.1$\times 10^{-4}$ &0.0 &0.0 &0.0 &0.0 &3.9$\times 10^{-3}$ &5.4$\times 10^{-3}$ &1.4$\times 10^{-5}$ \\
Oulu (103 subjects)       &0.0& 0.0&0.0 &0.0 &0.0 &0.0 &1.4$\times 10^{-4}$ &3.2$\times 10^{-4}$&0.0\\
\hline
\multicolumn{10}{|c|}{B2}\\
\hline
 &\multicolumn{3}{c|}{FEST}&\multicolumn{3}{c|}{FSTS}& \multicolumn{3}{c|}{FFBS}\\
\cline{2-10}
Source fMRI data     &   Marginal  & LTT & Joint&  Marginal & LTT & Joint&  Marginal  & LTT & Joint\\
\hline
Beijing (198 subjects)    &3.1$\times 10^{-6}$ &5.7$\times 10^{-6}$ &0.0 & 0.0& 0.0 &0.0 &7.0$\times 10^{-3}$ &1.2$\times 10^{-2}$& 1.3$\times 10^{-4}$\\
Cambridge (198 subjects)  &1.4$\times 10^{-3}$ &2.1$\times 10^{-3}$ &2.0$\times 10^{-5}$ &6.70$\times 10^{-8}$ &1.53$\times 10^{-5}$ & 2.9$\times 10^{-5}$&9.9$\times 10^{-3}$&1.4$\times 10^{-2}$ & 2.1$\times 10^{-4}$\\
Oulu (103 subjects)       &0.0&0.0 &0.0 &0.0 &0.0 &0.0 &1.2$\times 10^{-2}$ &2.1$\times 10^{-2}$&4.9$\times 10^{-4}$\\
\hline
\end{tabular}
\end{tiny}
\end{center}
\end{table}

In the table \ref{tab:table1}, we only show the assessment results for the variables B1 and B2 given that E1 and E2 yield pretty similar outcomes. Those results are congruent with what is observed in the assessmente with simulated data and real data examples shown above. The three algorithms present a low rate of false-positive activations, but with a less conservative behavior in the case of FEST and FFBS allowing for the detection of weaker BOLD signals. From these assessments (using real and simulated data), the examples presented above and other analysis that are not possible to be presented here, we would recommend the FEST as the best option among the three algorithms, not only because its good performance related to weak BOLD signals and false-positive activations, but also because its time of execution is 50\% faster in comparison with the FFBS and FSTS algorithms. 

\section{Concluding remarks}
\label{sec:conc}

In this work we develop an effective and reliable method for statistical analysis of fMRI data, taking into account its temporal and (at least locally) spatial features, without the need for the more traditional and complex spatial-temporal models. We employ an MDLM, which has shown to be a very flexible type of modeling beacuse of the availability of closed-form posterior distributions, discount factors to deal with the variability of the state parameter and the possibility to combine the posterior distributions through subjects to perform group analysis. Another important aspect of the MDLM is its relationship with the traditional and broadly known GLM (\cite[page 80]{west1997bayesian}), which can help to non-statistician practitioners to get more familiar with this type of modeling. We also make an important contribution, not only in the context of fMRI modeling but a general way to perform an assessment of dynamic effects on Dynamic Linear Models. From our knowledge, this is something that has been neglected, as it is mentioned above, possibly because of the type of applications and motivations behind MDLM. In this work, instead of making predictions or analyze covariance structures among time series, our aim is to assess affects size related to neural activity. Thus in order to achieve that using this type of modeling, we propose two new algorithms (FEST, FSTS) and adapt an existing one (FBST) to draw on-line trajectories related to the state parameter. Throughout the examples and assessments presented above, one can see that the performance of our method is remarkable, in the sense that it is able to both properly identify brain activation under different conditions of experimentation and control the rate of false activations. Currently we are working on the implementation of an R package for this method, which will be freely available for interested users from the fMRI community. It is worth mentioning that with this work we leave some open problems. One of them is the choice of the discount factor $\beta$, which from our experience dealing with fMRI data from different types of experiments just setting values for $0.8<\beta<1$ can work well. But instead of setting or tuning a particular value for $\beta$, a model selection approach for different $\beta$ values could be easily implemented in order to select the "best" $\beta$ value according to some statistical criteria such as Bayes Factors. Another problem is related to the distance parameter $r$, for which we believe a sensitivity analysis could be performed in order to have a better sense of the behavior of this relevant parameter. In all of the examples and analysis performed in this work, we set $r=1$. Aditionally, in the case of the FEST algorithm, there are some outputs such as the sequence of covariance matrices $\{\mathbf{S}_{v1}^{(z,k)}, \ldots, \mathbf{S}_{vt}^{(z,k)}\}$, which could be used to build some sort of dynamic graphs to tackle the connectivity problem among brain regions for resting-state fMRI data (see \cite{shen2010graph} for a better understanding of this type of analysis). Finally, we want to highlight that in this work only a tiny part of the rich possibilities offered by the general theory of the DLM's is explored. Our ideas could be extended to more complex model specifications or even to the case of non-linear Dynamic models.


\begin{description}

\item

\end{description}

\bibliographystyle{unsrtnat}
\bibliography{biblio}

\end{document}